\documentclass[12pt,dvips]{article}
\usepackage{rotating}
\usepackage{axodraw}
\usepackage{epsfig}
\usepackage{color}
\usepackage{cite}
\definecolor{GreenYellow}   {cmyk}{0.15,0,0.69,0}
\definecolor{Yellow}        {cmyk}{0,0,1,0}
\definecolor{Goldenrod}     {cmyk}{0,0.10,0.84,0}
\definecolor{Dandelion}     {cmyk}{0,0.29,0.84,0}
\definecolor{Apricot}       {cmyk}{0,0.32,0.52,0}
\definecolor{Peach}         {cmyk}{0,0.50,0.70,0}
\definecolor{Melon}         {cmyk}{0,0.46,0.50,0}
\definecolor{YellowOrange}  {cmyk}{0,0.42,1,0}
\definecolor{Orange}        {cmyk}{0,0.61,0.87,0}
\definecolor{BurntOrange}   {cmyk}{0,0.51,1,0}
\definecolor{Bittersweet}   {cmyk}{0,0.75,1,0.24}
\definecolor{RedOrange}     {cmyk}{0,0.77,0.87,0}
\definecolor{Mahogany}      {cmyk}{0,0.85,0.87,0.35}
\definecolor{Maroon}        {cmyk}{0,0.87,0.68,0.32}
\definecolor{BrickRed}      {cmyk}{0,0.89,0.94,0.28}
\definecolor{Red}           {cmyk}{0,1,1,0}
\definecolor{OrangeRed}     {cmyk}{0,1,0.50,0}
\definecolor{RubineRed}     {cmyk}{0,1,0.13,0}
\definecolor{WildStrawberry}{cmyk}{0,0.96,0.39,0}
\definecolor{Salmon}        {cmyk}{0,0.53,0.38,0}
\definecolor{CarnationPink} {cmyk}{0,0.63,0,0}
\definecolor{Magenta}       {cmyk}{0,1,0,0}
\definecolor{VioletRed}     {cmyk}{0,0.81,0,0}
\definecolor{Rhodamine}     {cmyk}{0,0.82,0,0}
\definecolor{Mulberry}      {cmyk}{0.34,0.90,0,0.02}
\definecolor{RedViolet}     {cmyk}{0.07,0.90,0,0.34}
\definecolor{Fuchsia}       {cmyk}{0.47,0.91,0,0.08}
\definecolor{Lavender}      {cmyk}{0,0.48,0,0}
\definecolor{Thistle}       {cmyk}{0.12,0.59,0,0}
\definecolor{Orchid}        {cmyk}{0.32,0.64,0,0}
\definecolor{DarkOrchid}    {cmyk}{0.40,0.80,0.20,0}
\definecolor{Purple}        {cmyk}{0.45,0.86,0,0}
\definecolor{Plum}          {cmyk}{0.50,1,0,0}
\definecolor{Violet}        {cmyk}{0.79,0.88,0,0}
\definecolor{RoyalPurple}   {cmyk}{0.75,0.90,0,0}
\definecolor{BlueViolet}    {cmyk}{0.86,0.91,0,0.04}
\definecolor{Periwinkle}    {cmyk}{0.57,0.55,0,0}
\definecolor{CadetBlue}     {cmyk}{0.62,0.57,0.23,0}
\definecolor{CornflowerBlue}{cmyk}{0.65,0.13,0,0}
\definecolor{MidnightBlue}  {cmyk}{0.98,0.13,0,0.43}
\definecolor{NavyBlue}      {cmyk}{0.94,0.54,0,0}
\definecolor{RoyalBlue}     {cmyk}{1,0.50,0,0}
\definecolor{Blue}          {cmyk}{1,1,0,0}
\definecolor{Cerulean}      {cmyk}{0.94,0.11,0,0}
\definecolor{Cyan}          {cmyk}{1,0,0,0}
\definecolor{ProcessBlue}   {cmyk}{0.96,0,0,0}
\definecolor{SkyBlue}       {cmyk}{0.62,0,0.12,0}
\definecolor{Turquoise}     {cmyk}{0.85,0,0.20,0}
\definecolor{TealBlue}      {cmyk}{0.86,0,0.34,0.02}
\definecolor{Aquamarine}    {cmyk}{0.82,0,0.30,0}
\definecolor{BlueGreen}     {cmyk}{0.85,0,0.33,0}
\definecolor{Emerald}       {cmyk}{1,0,0.50,0}
\definecolor{JungleGreen}   {cmyk}{0.99,0,0.52,0}
\definecolor{SeaGreen}      {cmyk}{0.69,0,0.50,0}
\definecolor{Green}         {cmyk}{1,0,1,0}
\definecolor{ForestGreen}   {cmyk}{0.91,0,0.88,0.12}
\definecolor{PineGreen}     {cmyk}{0.92,0,0.59,0.25}
\definecolor{LimeGreen}     {cmyk}{0.50,0,1,0}
\definecolor{YellowGreen}   {cmyk}{0.44,0,0.74,0}
\definecolor{SpringGreen}   {cmyk}{0.26,0,0.76,0}
\definecolor{OliveGreen}    {cmyk}{0.64,0,0.95,0.40}
\definecolor{RawSienna}     {cmyk}{0,0.72,1,0.45}
\definecolor{Sepia}         {cmyk}{0,0.83,1,0.70}
\definecolor{Brown}         {cmyk}{0,0.81,1,0.60}
\definecolor{Tan}           {cmyk}{0.14,0.42,0.56,0}
\definecolor{Gray}          {cmyk}{0,0,0,0.50}
\definecolor{Black}         {cmyk}{0,0,0,1}
\definecolor{White}         {cmyk}{0,0,0,0}

\newcommand{\imag}{\Im {\rm m}}
\newcommand{\real}{\Re {\rm e}}

\newcommand{\lsim}{\raisebox{-0.13cm}{~\shortstack{$<$ \\[-0.07cm] $\sim$}}~}
\newcommand{\gsim}{\raisebox{-0.13cm}{~\shortstack{$>$ \\[-0.07cm] $\sim$}}~}

\def\slash#1{#1\!\!\!/}

\begin{document}

\def\thefootnote{\fnsymbol{footnote}}

\begin{flushright}
NCU-HEP-k033\\
{\tt arXiv:0904.4352 [hep-ph]}\\
April 2009
\end{flushright}

\begin{center}
{\bf {\LARGE
Electric and anomalous magnetic dipole \\[3mm]
moments of the muon in the MSSM
} }
\end{center}

\medskip

\begin{center}{\large
Kingman Cheung$^{a,b,c}$,
Otto~C.~W.~Kong$^d$, and
Jae~Sik~Lee$^b$ }
\end{center}

\begin{center}
{\em $^a$ Department of Physics, National Tsing Hua University, Hsinchu, Taiwan 300}\\[0.2cm]
{\em $^b$Physics Division, National Center for Theoretical Sciences,
Hsinchu, Taiwan 300}\\[0.2cm]
{\em $^c$Division of Quantum Phases \& Devices,
                  Konkuk University,  Seoul 143-701, Korea}\\[0.2cm]
{\em $^d$Department of Physics and Center for Mathematics and
  Theoretical Physics,}\\
{\em National Central University, Chung-Li, Taiwan 32054}
\end{center}

\bigskip\bigskip

\centerline{\bf ABSTRACT}
\begin{flushleft}
\medskip\noindent  
We study the electric dipole moment (EDM) and the anomalous 
magnetic dipole moment (MDM) of the muon in the 
CP-violating Minimal
Supersymmetric extension  of the Standard Model~(MSSM).   
We take into account the contributions from 
the chargino- and neutralino-mediated one-loop graphs
and the dominant two-loop Higgs-mediated Barr-Zee diagrams.
We improve earlier calculations by incorporating
CP-violating Higgs-boson mixing effects and the
resummed threshold corrections to the Yukawa couplings of 
the charged leptons as well as 
that of the bottom quark.
The analytic correlation between the muon EDM and MDM is explicitly
presented at one- and two-loop levels and,
through several numerical examples, we illustrate its dependence 
on the source of the dominant contributions.
We have implemented the analytic expressions for the muon EDM and MDM in
an updated version of the public code {\tt CPsuperH2.0}.
\end{flushleft}

\newpage 
\section{Introduction}
The anomalous magnetic dipole moment of the muon, $a_\mu\,$,
has provided one of the most sensitive test grounds for 
the validity of the Standard Model (SM)~\cite{ref:exp}:
\begin{equation}
a_\mu^{\rm EXP} = 11\,659\,208\,(6.3)\,\times 10^{-10}\,.
\label{eq:EXP}
\end{equation}
At the same time, it also provides an important constraint on 
new physics with precise SM predictions available. The SM prediction consists
of QED, electroweak (EW), and hadronic contributions. The
hadronic contribution is further decomposed into 
leading-order (LO) part and higher-order
vacuum polarization (VP) and
light-by-light (LBL) parts~\cite{SM:REVIEW,SM:QED1,SM:QED2,SM:QED3,
SM:EW,SM:HLO1,SM:HLO2,SM:HLO3,SM:HLO4,SM:HLO5,SM:HVP,
SM:HLBL1,SM:HLBL2,SM:HLBL3,SM:HLBL3.5,SM:HLBL4,SM:HLBL5}
\footnote{See Table~\ref{tab:SM} for details of the SM prediction.}:
\begin{eqnarray}
a_\mu^{\rm SM} &=& a_\mu^{\rm QED} + a_\mu^{\rm EW} + a_\mu^{\rm Had.(LO)}
+ a_\mu^{\rm Had.(VP)} + a_\mu^{\rm Had.(LBL)} \nonumber \\
&=& 11\,659\,177.3\,(5.3)\times 10^{-10}\,.
\label{eq:SM}
\end{eqnarray}
Equations~(\ref{eq:EXP}) and (\ref{eq:SM}) suggest that
there is currently a $3.7\,\sigma$ discrepancy between
the experimental result and the SM prediction,
 which can be attributed to possible
contributions from physics beyond the SM
\footnote{
The number incorporates the hadronic VP result calculated based on
measurements at electron-position storage ring. Calculation based on
hadronic $\tau$-decays gives a different result, which is considered
less reliable. If the $\tau$-based result is used, the overall
discrepancy reduces to only about 1$\sigma$.}:
\begin{equation}
\Delta a_\mu^{\rm EXP} \equiv 
a_\mu^{\rm EXP}-a_\mu^{\rm SM} = 30.7\,(8.2)\times 10^{-10}~(3.7\,\sigma)\,.
\label{eq:damu}
\end{equation}
 
One of the most appealing scenarios for physics beyond the SM
is augmented with a softly broken supersymmetry (SUSY) around the TeV scale. The 
supersymmetric contributions to $a_\mu$ from such models are known up to 
dominant two-loop contributions. 
The one-loop results can be found in
\cite{one_loop_MDM:earlier,one_loop_MDM:later1,one_loop_MDM:later2,
one_loop_MDM:recent1,one_loop_MDM:recent2,one_loop_MDM:recent2.5}
and  the two-loop results in
\cite{Heinemeyer:2003dq,Heinemeyer:2004yq,Degrassi:1998es,Feng:2006ei,
Marchetti:2008hw,Feng:2008cn}.
It is well-known that the dominant two-loop contribution comes from 
Higgs-mediated Barr-Zee diagrams~\cite{Barr-Zee_MDM}.
The error associated with the known SUSY contributions is estimated to be
$\sim 2.5 \times 10^{-10}$~\cite{Stockinger:2006zn}, which is 
smaller than half of the
current experimental and SM theoretical ones.
%
\begin{table}[\hbt]
\label{tab:SM}
\caption{The SM prediction of $a_\mu^{\rm SM}$: see, for example,
Ref.~\cite{SM:REVIEW}. For the quoted value in Eq.(\ref{eq:SM})
we use the results in \cite{SM:HLO2} and \cite{SM:HLBL4} for the hadronic leading-order
and the hadronic light-by-light contributions, respectively.}
\begin{center}
\begin{tabular}{|c|r|l|l|}
\hline
& $a_\mu \times 10^{10}~~~$ & $~\delta a_\mu \times 10^{10}$ & Ref.\\
\hline \hline
$a_\mu^{\rm QED}$& $11\,658\,471.810$ & $(0.016)$ &
\protect\cite{SM:QED1,SM:QED2,SM:QED3}\\
\hline
$a_\mu^{\rm EW}$& $15.4$ & $(0.2)$ &        \protect\cite{SM:EW}\\
\hline
$a_\mu^{\rm Had.(LO)}$& $690.9$ & $(4.4)$&  \protect\cite{SM:HLO1}\\
$a_\mu^{\rm Had.(LO)}$& $689.4$ & $(4.6)$&  \protect\cite{SM:HLO2} $(*)$\\
$a_\mu^{\rm Had.(LO)}$& $692.1$ & $(5.6)$&  \protect\cite{SM:HLO3}\\
$a_\mu^{\rm Had.(LO)}$& $694.4$ & $(4.9)$&  \protect\cite{SM:HLO4}\\
$a_\mu^{\rm Had.(LO)}$& $691.04$ & $(5.29)$&  \protect\cite{SM:HLO5}\\
\hline
$a_\mu^{\rm Had.(VP)}$& $-9.8$ & ($0.1)$&   \protect\cite{SM:HLO2,SM:HVP}\\
\hline
$a_\mu^{\rm Had.(LBL)}$& $8.0$ & $(4.0)$&   \protect\cite{SM:HLBL1}\\
$a_\mu^{\rm Had.(LBL)}$& $13.6$ & $(2.5)$&  \protect\cite{SM:HLBL2}\\
$a_\mu^{\rm Had.(LBL)}$& $11.0$ & $(4.0)$&  \protect\cite{SM:HLBL3}\\
$a_\mu^{\rm Had.(LBL)}$& $11.6$ & $(4.0)$&  \protect\cite{SM:HLBL3.5}\\
$a_\mu^{\rm Had.(LBL)}$& $10.5$ & $(2.6)$&  \protect\cite{SM:HLBL4} $(*)$\\
$a_\mu^{\rm Had.(LBL)}$& $10.2$ & $({\tiny\sim} 3)$&  \protect\cite{SM:HLBL5}\\
\hline
\end{tabular}
\end{center}
\end{table}
%
On the other hand, the SUSY augmented models can contain additional CP-violating
phases beyond the SM Cabibbo-Kobayashi-Maskawa (CKM) phase
leading to sizable EDMs
\cite{Pospelov:2005pr,RamseyMusolf:2006vr,Ibrahim:2007fb,Ellis:2008zy}.
The current limit on the muon EDM~\cite{Bennett:2008dy} is
\begin{equation}
|d_\mu| < 1.8 \times 10^{-19}~{\rm e\,cm}~(95 \% \, {\rm C.L.})\,,
\end{equation}
which is much weaker than the constraints from the non-observation of the
Thallium~\cite{Regan:2002ta}, neutron~\cite{Baker:2006ts},  and
Mercury~\cite{Romalis:2000mg} EDMs:
\begin{eqnarray}
|d_{\rm Tl}| &<& 9\times 10^{-25}~e\,{\rm cm}\,, \nonumber \\
|d_{\rm n}|  &<& 3\times 10^{-26}~e\,{\rm cm}\,, \nonumber \\
|d_{\rm Hg}| &<& 2\times 10^{-28}~e\,{\rm cm}\,. 
\end{eqnarray}
Nevertheless, if the muon EDM experiment in the future can achieve 
the projected sensitivity~\cite{Semertzidis:1999kv}
\begin{equation}
d_\mu \sim 1 \times 10^{-24} ~e\,{\rm cm}\,,
\end{equation}
the precision of the experiment will be comparable to 
that of the current Thallium EDM experiment.

In this paper, we study the correlation between the muon EDM and 
MDM in the CP-violating MSSM~\cite{Ibrahim:2001ym,Feng:2008cn}.
We present the relation between the one-loop
chargino- and neutralino-mediated EDM and MDM of the muon. 
We also derive an analytic relation between them in the two-loop 
contributions from the dominant Higgs-mediated Barr-Zee diagrams.
We improve the earlier results by including
CP-violating Higgs-boson mixing effects in the Barr-Zee diagrams
and resumming the threshold corrections to the muon, tau, and bottom-quark
Yukawa couplings in the one- and two-loop graphs.
We then focus our numerical studies on three types of
scenarios in which $(i)$ the muon EDM and MDM are dominated by the
one-loop contributions, $(ii)$ the lightest Higgs
boson is mostly CP odd and
lighter than $\sim$ 50 GeV, and $(iii)$ the dominant contributions 
to the muon EDM and MDM come from the two-loop Barr-Zee graphs.

For  the   presentation  of  our  analytic  results,   we  follow  the
conventions   and   notations   of   {\tt   CPsuperH}~\cite{cpsuperh},
especially for  the masses  and mixing matrices  of the  neutral Higgs
bosons  and SUSY  particles. 
The layout of the paper is as follows. Section~2 presents formulas relevant
to the one-loop contributions to the muon EDM and MDM from  chargino- and
neutralino-mediated diagrams. Non-holomorphic threshold effects on the 
muon Yukawa coupling have been appropriately  resummed. In Section~3, we  
present analytic results for the Higgs-mediated
two-loop Barr-Zee diagrams.  For this, most importantly,
the resummed threshold corrections to
the tau and bottom-quark Yukawa couplings and
the CP-violating Higgs-boson mixing effects have been incorporated.
In Section~4, we present some numerical examples,
depending on the source of dominant contributions to the muon EDM and MDM.
We summarize our findings in Section~5.

%

\section{One-Loop EDMs and MDMs of charged leptons}
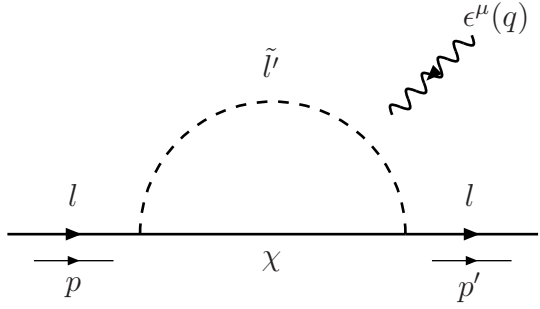
\begin{figure}[t]
\begin{center}
\begin{picture}(250,100)(0,0)
\SetWidth{1.0}
\ArrowLine(25,25)(75,25)
\Line(75,25)(175,25)
\ArrowLine(175,25)(225,25)
\DashCArc(125,25)(50,0,180){4}
\Photon(170,70)(200,100){3}{5}
\DashArrowLine(186,86)(184,84){1}

\SetWidth{0.5}
\ArrowLine(35,15)(65,15)\Text(50,5)[]{$p$}\Text(50,40)[]{$l$}
\ArrowLine(185,15)(215,15)\Text(200,5)[]{$p^\prime$}\Text(200,40)[]{$l$}
\Text(125,15)[]{$\chi$}
\Text(125,90)[]{$\tilde{l'}$}
\Text(210,107)[]{$\epsilon^\mu (q)$}
\end{picture} 
\end{center}
\smallskip
\noindent
\caption{\it Feynman diagrams for the EDM $(d_l)^\chi$ and
MDM $(a_l)^\chi$ of the lepton $l$
induced by $\chi$. The photon line can be attached
to the slepton $\tilde{l'}$ line
or the $\chi$ line.}
\label{fig:EMDM}
\end{figure}

The relevant interaction Lagrangian of the spin-1/2 lepton with EDM $d_l$ and MDM $a_l$ 
is given by
\begin{equation}
{\cal L}_{\rm
spin-1/2}=-\frac{i}{2}\,d_l\,(\bar{l}\,\sigma_{\mu\nu}\,\gamma_5\,l)\,F^{\mu\nu}
+\frac{e\,a_l}{4\,m_l}\,(\bar{l}\,\sigma_{\mu\nu}\,l)\,F^{\mu\nu}\,,
\label{eq:MDM_EDM}
\end{equation}
where $\sigma^{\mu\nu}=
\frac{i}{2}[\gamma^\mu\,,\gamma^\nu]=i(\gamma^\mu\gamma^\nu -g^{\mu\nu})$. 
The EDM and MDM amplitudes are given by
\begin{equation}
({\cal M}_{\rm MDM})^\mu=\frac{e\, a_l}{2 m_l}\,\bar{u}(p') (i \sigma^{\mu\nu} q_\nu)
u(p)\,,
\ \ \
({\cal M}_{\rm EDM})^\mu=d_l\,\bar{u}(p') (\sigma^{\mu\nu} \gamma_5 q_\nu) u(p)\,.
\end{equation} 
 Generic interactions of charginos $\tilde{\chi}^\pm_{1,2}$ or
neutralinos $\tilde{\chi}^0_{1,2,3,4}$, collectively denoted by $\chi$, 
with a lepton $l$ and a slepton (sneutrino) 
$\tilde{l}_{1,2}^\prime$ ($\tilde\nu_l$) are given by
\begin{equation}
{\cal L}_{\chi l \tilde{l}^\prime} = g_{L\,ij}^{\chi l \tilde{l}^\prime}
(\bar{\chi_i} P_L l)\, \tilde{l}_i^{\prime *}
+ g_{R\,ij}^{\chi l \tilde{l}^\prime}\, (\bar{\chi_i} P_R l)\, \tilde{l}_j^{\prime *} 
+ {\rm h.c.}
\end{equation}
The Lagrangian for the interactions of $\chi$ and $\tilde{l}_{1,2}^\prime$ 
with the photon field $A_\mu$ is 
\begin{equation}
{\cal L}_{\chi\chi A}=-e\,Q_\chi\,(\bar{\chi}\gamma_\mu \chi) A^\mu \ \ {\rm and} \ \
{\cal L}_{\tilde{l'}\tilde{l'}A}=-ie\,Q_{\tilde{l'}}\,\tilde{l'}^*
\stackrel {\leftrightarrow} {\partial}_\mu \tilde{l'} A^\mu\,.
\end{equation}
The diagrams in Fig.~\ref{fig:EMDM} induce the MDM and EDM of the 
lepton $l$ as follows:
\begin{eqnarray}
\left(a_l\right)^\chi&=&\frac{m_l^2}{8\pi^2 m_{\tilde{l}'_j}^2}\Bigg\{
\left(\left|g_{R\,ij}^{\chi l \tilde{l}'}\right|^2
+\left|g_{L\,ij}^{\chi l \tilde{l}'}\right|^2\right)
\left[-Q_\chi\, {\cal A}(m_{\chi_i}^2/m_{\tilde{l}'_j}^2)
+Q_{\tilde{l'}}\, {\cal B}(m_{\chi_i}^2/m_{\tilde{l}'_j}^2)\right]
\nonumber \\
&&~~\,+\frac{m_{\chi_i}}{m_l}\,
\real\left[\left(g_{R\,ij}^{\chi l \tilde{l}'}\right)^*\,g_{L\,ij}^{\chi l \tilde{l}'}\right]
\left[Q_\chi\, A(m_{\chi_i}^2/m_{\tilde{l}'_j}^2)
+Q_{\tilde{l'}}\, B(m_{\chi_i}^2/m_{\tilde{l}'_j}^2)\right]\Bigg\}\,,
\nonumber \\
\left(\frac{d_l}{e}\right)^\chi&=&\frac{m_{\chi_i}}{16\pi^2 m_{\tilde{l}'_j}^2}
\imag\left[\left(g_{R\,ij}^{\chi l \tilde{l}'}\right)^*\,g_{L\,ij}^{\chi l \tilde{l}'}\right]
\left[Q_\chi\, A(m_{\chi_i}^2/m_{\tilde{l}'_j}^2)
+Q_{\tilde{l'}}\, B(m_{\chi_i}^2/m_{\tilde{l}'_j}^2)\right]\,,
\label{eq:oneloop}
\end{eqnarray}
where
\begin{eqnarray}
A(r)&=& \frac{1}{2(1-r)^2}\left(3-r+\frac{2 \ln{r}}{1-r}\right)\,, \nonumber \\
B(r)&=& \frac{1}{2(1-r)^2}\left(1+r+\frac{2 r \ln{r}}{1-r}\right)\,, \nonumber \\
{\cal A}(r)&=& \frac{1}{12(1-r)^3}\left(2+5r-r^2+\frac{6r \ln{r}}{1-r}\right)\,, \nonumber \\
{\cal B}(r)&=& \frac{1}{12(1-r)^3}\left(1-5r-2r^2-\frac{6r^2\ln{r}}{1-r}\right)\,.
\end{eqnarray}
We have checked that our analytic expressions for the one-loop MDM and EDM
agree with those given in, for example, 
Ref.~\cite{one_loop_MDM:recent2} and \cite{Ellis:2008zy}
with $Q_{\tilde\chi_i^-}=Q_{\tilde{l}_j^-}=-1$ and
$Q_{\tilde\chi_k^0}=Q_{\tilde{\nu}_l}=0$.
Note $A(1)=-1/3$, $B(1)=1/6$ and ${\cal A}(1)={\cal B}(1)=1/24$.
Finally, in Eq.~(\ref{eq:oneloop}),
the chargino-lepton-sneutrino couplings are given by
\begin{eqnarray}
g_{L\,i}^{\tilde{\chi}^\pm l \tilde{\nu}} = -g (C_R)_{i1}\,, \ \ \ \
g_{R\,i}^{\tilde{\chi}^\pm l \tilde{\nu}} = h_l^* (C_L)_{i2}\,,
\end{eqnarray}
and the neutralino-lepton-slepton couplings are given by
\begin{eqnarray}
g_{L\,ij}^{\tilde{\chi}^0 l \tilde{l}} &=& 
-\sqrt{2}\, g\, T_3^l\, N_{i2}^* (U^{\tilde{l}})_{1j}^*
-\sqrt{2}\, g\, t_W\, (Q_l-T_3^l) N_{i1}^* (U^{\tilde{l}})_{1j}^*
-h_l N_{i3}^* (U^{\tilde{l}})_{2j}^*\,, \nonumber \\
g_{R\,ij}^{\tilde{\chi}^0 l \tilde{l}} &=&
\sqrt{2}\, g\, t_W\, Q_l\, N_{i1} (U^{\tilde{l}})_{2j}^*
-h_l^* N_{i3} (U^{\tilde{l}})_{1j}^*\,, 
\end{eqnarray}
with $T_3^{l}=-1/2$ and $Q_l=-1$.

\vspace*{0.5cm}

There are non-holomorphic threshold corrections to the
Yukawa couplings $h_l$ which appear in the chargino and neutralino couplings
\cite{Guasch:2001wv,Carena:1998gk,Ibrahim:2003tq}.
These corrections become significant at large $\tan\beta$ and we resum
these effects by redefining
the Yukawa couplings as follows:
\begin{equation}
h_l = \frac{\sqrt{2} \,m_l}{v\, c_\beta}\,\frac{1}{1+\Delta_l\,t_\beta}\,.
\end{equation}
In the presence of the CP phases, $\Delta_l$ takes the form \cite{Ellis:2004fs}
\begin{eqnarray}
\Delta_l &=&
-\, \frac{\alpha_{\rm em}\,\mu^*M_2^*}{4\pi\,s^2_W}\,
\Bigg[\, I(m_{\tilde{\nu}_l}^2,|M_2|^2,|\mu|^2)\:
+\: \frac{1}{2}\, |U^{\tilde{l}}_{L1}|^2\,
I(m_{\tilde{l}_1}^2,|M_2|^2,|\mu|^2)\nonumber\\
&&+\, \frac{1}{2}\, |U^{\tilde{l}}_{L2}|^2\,
I(m_{\tilde{l}_2}^2,|M_2|^2,|\mu|^2)\,\Bigg]
\ +\ \frac{\alpha_{\rm em}\,\mu^*M_1^*}{4\pi\,c_W^2}\,
\Bigg[\, I(m_{\tilde{l}_1}^2,m_{\tilde{l}_2}^2,
|M_1|^2)\nonumber\\
&& +\: \frac{1}{2}\, |U^{\tilde{l}}_{L1}|^2\,
I(m_{\tilde{l}_1}^2,|M_1|^2,|\mu|^2)\: +\:
\frac{1}{2}\,
|U^{\tilde{l}}_{L2}|^2\, I(m_{\tilde{l}_2}^2,|M_1|^2,|\mu|^2)
\nonumber \\
&&-\,|U^{\tilde{l}}_{R1}|^2\,I(m_{\tilde{l}_1}^2,|M_1|^2,|\mu|^2)
\: -\: |U^{\tilde{l}}_{R2}|^2\,
I(m_{\tilde{l}_2}^2,|M_1|^2,|\mu|^2)\Bigg]\,,
\end{eqnarray}
where the one-loop function is
\begin{equation}
  \label{Ixyz}
I(x,y,z)\  \equiv \ \frac{xy\,\ln (x/y)\: +\: yz\,\ln (y/z)\: +\:
             xz\, \ln (z/x)}{(x-y)\,(y-z)\,(x-z)}\,.
\end{equation}
These effects on the muon MDM
have been considered in Ref.~\cite{Marchetti:2008hw} only in the CP conserving case.
\section{Barr-Zee Graphs}
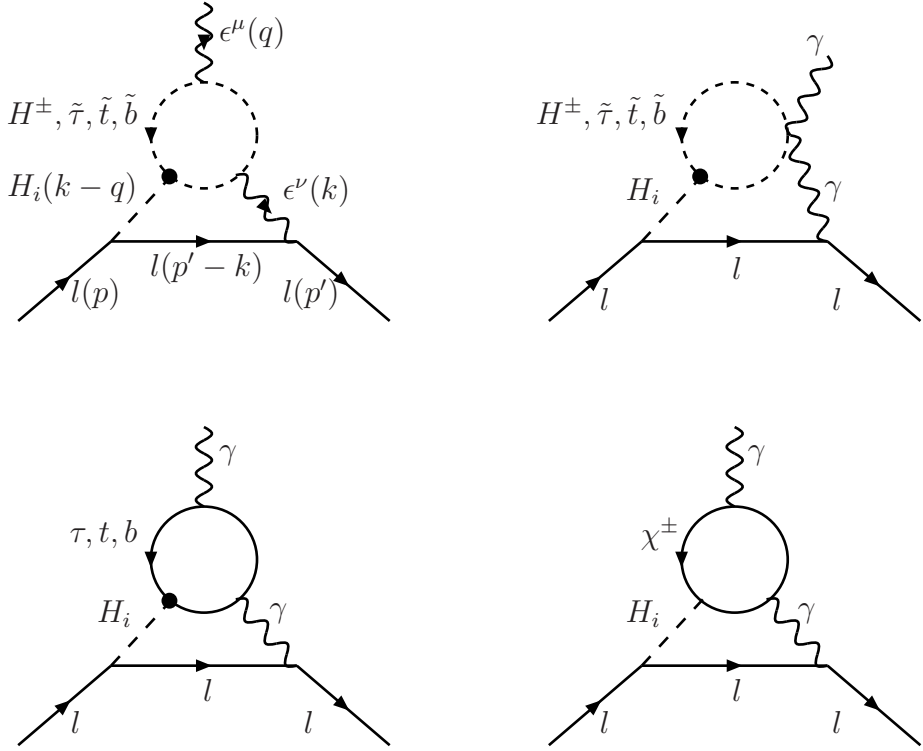
\begin{figure}[t]
\begin{center}
\begin{picture}(400,300)(0,70)
\SetWidth{1.0}
\ArrowLine(10,230)(45,260)\Text(30,235)[lb]{$l(p)$}
\DashLine(45,260)(68,285){5}\Vertex(67,284.5){3}
\Text(55,275)[rb]{$H_i(k-q)$}
\Photon(92,285)(115,260){3}{3}\Text(110,280)[l]{$\epsilon^\nu(k)$}
\DashArrowLine(103,273)(104,272){1}
\ArrowLine(45,260)(115,260)\Text(60,251)[l]{$l (p'-k)$}
\ArrowLine(115,260)(150,230)\Text(110,235)[lb]{$l (p')$}
\Photon(80,320)(80,350){3}{3}\Text(86,340)[l]{$\epsilon^\mu(q)$}
\DashArrowLine(80,336)(80,334){1}
\DashArrowArc(80,300)(20,0,360){3}
\Text(55,310)[r]{$H^\pm,\tilde{\tau},\tilde{t},\tilde{b}$}


\ArrowLine(210,230)(245,260)\Text(230,235)[lb]{$l$}
\DashLine(245,260)(268,285){5}\Vertex(267,284.5){3}
\Text(253,275)[rb]{$H_i$}
\Photon(301,300)(315,260){3}{4}\Text(315,280)[l]{$\gamma$}
\Photon(301,300)(315,330){3}{3}\Text(315,335)[r]{$\gamma$}
\ArrowLine(245,260)(315,260)\Text(280,251)[l]{$l$}
\ArrowLine(315,260)(350,230)\Text(318,235)[lb]{$l$}
\DashArrowArc(280,300)(20,0,360){3}
\Text(255,310)[r]{$H^\pm,\tilde{\tau},\tilde{t},\tilde{b}$}


\ArrowLine(10,70)(45,100)\Text(30,75)[lb]{$l$}
\DashLine(45,100)(68,125){5}\Vertex(67,124.5){3}
\Text(53,115)[rb]{$H_i$}
\Photon(92,125)(115,100){3}{3}\Text(105,120)[l]{$\gamma$}
\ArrowLine(45,100)(115,100)\Text(80,91)[l]{$l$}
\ArrowLine(115,100)(150,70)\Text(118,75)[lb]{$l$}
\Photon(80,160)(80,190){3}{3}\Text(86,180)[l]{$\gamma$}
\ArrowArc(80,140)(20,0,360)
\Text(55,150)[r]{$\tau,t,b$}

\ArrowLine(210,70)(245,100)\Text(230,75)[lb]{$l$}
\DashLine(245,100)(268,125){5}
\Text(253,115)[rb]{$H_i$}
\Photon(292,125)(315,100){3}{3}\Text(305,120)[l]{$\gamma$}
\ArrowLine(245,100)(315,100)\Text(280,91)[l]{$l$}
\ArrowLine(315,100)(350,70)\Text(318,75)[lb]{$l$}
\Photon(280,160)(280,190){3}{3}\Text(286,180)[l]{$\gamma$}
\ArrowArc(280,140)(20,0,360)
\Text(260,150)[r]{$\chi^\pm$}

\end{picture}
\end{center}
\smallskip
\noindent
\caption{\it  Barr-Zee  diagrams: the  $H_i$  lines  denote all  three
neutral Higgs bosons,  including CP-violating Higgs-boson mixing, and heavy
dots   indicate   resummation   of   threshold  corrections   to   the
corresponding Yukawa couplings.}
\label{fig:Barr-Zee}
\end{figure}

The dominant two-loop contributions come from the 
Barr-Zee diagrams mediated by neutral Higgs-boson exchanges, see Fig.~\ref{fig:Barr-Zee}.
Here, we consider
loops of third-generation fermions and sfermions, charged Higgs bosons, 
and charginos.
The diagrams can be evaluated first by one-loop computation of the
$H_i(k-q)$--$\gamma^\mu(q)$--$\gamma^\nu(k)$ vertex. 
Thanks to gauge invariance
the effective vertex takes the form
\begin{equation}
\Gamma^{\mu\nu}_i=\left[g^{\mu\nu}k\cdot q-k^\mu q^\nu\right]{\cal S}_i(k^2)
+\left[\epsilon^{\mu\nu\alpha\beta}k_\alpha q_\beta\right]{\cal P}_i(k^2)\,,
\end{equation}
where $q$ is the incoming four-momentum of the external photon, 
$p$ ($p'$) the four-momentum of the incoming (outgoing) lepton, and
$k$ the four-momentum of the internal photon going out of the
upper loop. Note that $p'=p+q$. 
Explicitly, 
keeping only the terms linear in the external momenta,
the fermionic contributions are given by
\begin{eqnarray}
{\cal S}^f_i(k^2) &=& -\frac{1}{4\pi^2} N_C^f\, e^2 Q_f^2 g_f m_f\,g^S_{H_i\bar{f}f}\,
\int_0^1\frac{1-2x(1-x)}{x(1-x)k^2-m_f^2}\,,\nonumber \\
{\cal P}^f_i(k^2) &=& -\frac{1}{4\pi^2} N_C^f\, e^2 Q_f^2 g_f m_f\,g^P_{H_i\bar{f}f}\,
\int_0^1\frac{1}{x(1-x)k^2-m_f^2}\,,
\end{eqnarray}
with the generic $f$-$f$-$H_i$ interaction:
${\cal L}_{H_i\bar{f}f}=
-g_f\,\bar{f}\,(g^S_{H_i\bar{f}f}+i\gamma_5\,g^P_{H_i\bar{f}f})\,f\,H_i\,.$
The color factor
$N_C^f=3$ for quarks and 1 for leptons and charginos. 
On the other hand, the sfermion
loops contribute only to the scalar form factor as
\begin{equation}
{\cal S}^{\tilde{f}}_i(k^2) = \frac{1}{8\pi^2} 
N_C^f\, e^2 Q_f^2\, v \,g_{H_i\tilde{f}^*\tilde{f}}\,
\int_0^1\frac{x(1-x)}{x(1-x)k^2-m_{\tilde{f}}^2}\,,
\end{equation}
with the generic $\tilde{f}$-$\tilde{f}$-$H_i$ interaction:
${\cal L}_{H_i\tilde{f}^*\tilde{f}}=
v\,g_{H_i\tilde{f}^*\tilde{f}}\,\tilde{f}^*\tilde{f}\,H_i\,.$
%

Together with the interaction
Lagrangian for the couplings of the neutral-Higgs bosons and photon
to charged leptons
\begin{equation}
{\cal L}_{H_i\bar{l}l}=
-g_l\,\bar{l}\,(g^S_{H_i\bar{l}l}+i\gamma_5\,g^P_{H_i\bar{l}l})\,l\,H_i\,, 
\ \ \
{\cal L}_{A\bar{l}l}=-e\,Q_l\,\bar{l}\,\gamma^\mu\,l\,A_\mu\,,
\end{equation}
where  $g_l=\frac{g\, m_l}{2 M_W}$ (as in the {\tt CPsuperH} convention),
the Barr-Zee amplitude is given by
\begin{eqnarray}
i\,{\cal M}_{\rm Barr-Zee}^\mu=&&  \\
&&\hspace{-3.0cm}
e g_l Q_l\,\int\frac{{\rm d}^4k}{(2\pi)^4}
\frac{
\bar{u}(p')\left[
\Gamma^{\mu\nu}_i\gamma_\nu\,(-\slash{k})\,
(g^S_{H_i\bar{l}l}+i\gamma_5\,g^P_{H_i\bar{l}l})+
(g^S_{H_i\bar{l}l}+i\gamma_5\,g^P_{H_i\bar{l}l})\,(\slash{k})\,
\Gamma^{\mu\nu}_i\gamma_\nu\,\right] u(p)
}{k^2\,k^2\,(k^2-M_{H_i}^2)}\,, \nonumber
\end{eqnarray}
where again we keep only the terms linear in the external momenta. 
We note the numerator of the integrand is proportional to
\begin{equation}
(g^S_{H_i\bar{l}l}\,{\cal S}_i - g^P_{H_i\bar{l}l}\,{\cal P}_i)
(i\sigma^{\mu\nu} q_\nu)
-(g^P_{H_i\bar{l}l}\,{\cal S}_i + g^S_{H_i\bar{l}l}\,{\cal P}_i)
(\sigma^{\mu\nu} \gamma_5 q_\nu)\,.
\end{equation}
We observe that
the first term gives the MDM while the second one gives the EDM of the lepton $l$.
Consequently, the Barr-Zee contributions to the MDM and EDM are related by\\
\begin{equation}
\label{bbb}
(a_l)^H = 2 m_l \left.\left(\frac{d_l}{e}\right)^H\right|_
{\left\{\begin{array}{l}g^S_{H_i\bar{l}l} \to g^P_{H_i\bar{l}l} \\ g^P_{H_i\bar{l}l} \to -
g^S_{H_i\bar{l}l} \end{array}\right.}\,,
\end{equation}
with normalizations of the MDM and EDM of the lepton $l$ as given by
Eq.~(\ref{eq:MDM_EDM}). For the dipole moment diagrams with chirality flip
inside the loop, the MDM and EDM parts correspond directly to real and imaginary 
parts of the overall amplitude. The above relation is a recasting of that
statement in terms of the effective scalar and pseudoscalar couplings of the
generally CP-mixed Higgs states involved. To give the MDM result
explicitly, with $Q_l=-1$, we  have
\footnote{Here we add the contribution from the charged Higgs boson loop and
confirm the positive signs of the fermionic Barr-Zee
contributions~\cite{Ellis:2008zy}.}
\begin{eqnarray}
\left(a_l\right)^H &=& \sum_{q=t,b}\Bigg\{
-\frac{3\alpha_{\rm em}\,Q_q^2\,m_l^2}{16\pi^3}\sum_{i=1}^3\frac{g^S_{H_il^+l^-}}{M_{H_i}^2}
\sum_{j=1,2} g_{H_i\tilde{q}_j^*\tilde{q}_j}\,F(\tau_{\tilde{q}_ji})
\nonumber \\
&& +\frac{3\alpha_{\rm em}^2\,Q_q^2\,m_l^2}{4\pi^2s_W^2M_W^2}
\sum_{i=1}^3\left[
-g^S_{H_il^+l^-} g^S_{H_i\bar{q}q}\,f(\tau_{qi})
+g^P_{H_il^+l^-} g^P_{H_i\bar{q}q}\,g(\tau_{qi})
\right]\Bigg\}  \nonumber \\
&&-\frac{\alpha_{\rm em}\,m_l^2}{16\pi^3}\sum_{i=1}^3\frac{g^S_{H_il^+l^-}}{M_{H_i}^2}
\sum_{j=1,2} g_{H_i\tilde{\tau}_j^*\tilde{\tau}_j}\,F(\tau_{\tilde{\tau}_ji})
\nonumber \\
&&+\frac{\alpha_{\rm em}^2\,m_l^2}{4\pi^2s_W^2M_W^2}
\sum_{i=1}^3\left[
-g^S_{H_il^+l^-} g^S_{H_i\tau^+\tau^-}\,f(\tau_{\tau i})
+g^P_{H_il^+l^-} g^P_{H_i\tau^+\tau^-}\,g(\tau_{\tau i})
\right]\,, \nonumber \\
&&-\frac{\alpha_{\rm em}\,m_l^2}{16\pi^3}\sum_{i=1}^3\frac{g^S_{H_il^+l^-}}{M_{H_i}^2}
g_{H_iH^+H^-}\,F(\tau_{H^\pm i})
\nonumber \\
&&+\frac{\alpha_{\rm em}^2\,m_l^2}{2\sqrt{2}\pi^2s_W^2M_W} \nonumber \\
&&\hspace{0.5cm}\times
\sum_{i=1}^3\sum_{j=1,2}\frac{1}{m_{\chi^\pm_j}}\left[
-g^S_{H_il^+l^-} g^S_{H_i\chi^+_j\chi^-_j}\,f(\tau_{\chi_j^\pm i})
+g^P_{H_il^+l^-} g^P_{H_i\chi^+_j\chi^-_j}\,g(\tau_{\chi_j^\pm i})
\right]\,,
\label{eq:ah}
\end{eqnarray}
where $\tau_{xi}=m_x^2/M_{H_i}^2$ and
the two-loop functions $F(\tau )$,  $f(\tau )$, and $g(\tau )$ are
\begin{eqnarray}
F(\tau) &=& \int_0^{1} dx\ \frac{x(1-x)}{\tau\: -\: x(1-x)}\
\ln \bigg[\,\frac{x(1-x)}{\tau}\,\bigg]\, , \nonumber \\
f(\tau) &=& \frac{\tau}{2}\, \int_0^{1} dx\ \frac{1\: -\: 2x(1-x)}{x(1-x)\: -\: \tau}\
\ln \bigg[\,\frac{x(1-x)}{\tau}\,\bigg]\, , \nonumber \\
g(\tau) &=& \frac{\tau}{2}\, \int_0^{1} dx\ \frac{1}{x(1-x)\: -\: \tau}\
\ln \bigg[\,\frac{x(1-x)}{\tau}\,\bigg]\,.
\end{eqnarray}

For genuine SUSY contributions, the embedded SM contribution should be subtracted. 
We estimate the SM contributions as
\begin{eqnarray}
\left(a_l\right)^H_{\rm \,SM} = 
-\sum_{q=t,b}\left[
\frac{3\alpha_{\rm em}^2\,Q_q^2\,m_l^2}{4\pi^2s_W^2M_W^2}
\,f(\tau_{q{\rm SM}})\right]
-\frac{\alpha_{\rm em}^2\,m_l^2}{4\pi^2s_W^2M_W^2}
\,f(\tau_{\tau{\rm SM}}) \;,
\end{eqnarray}
where $\tau_{f{\rm SM}}=m_f^2/M_{H_{\rm SM}}^2$.
We find that $\left|\left(a_\mu\right)^H_{\rm \,SM}\right| 
\simeq 1.8 \times 10^{-11}$ when $M_{H_{\rm
SM}}=100$ GeV and it decreases as $M_{H_{\rm SM}}$ increases.
In our numerical analysis, we safely neglect $(a_l)^H_{\rm \,SM}$.
\section{Numerical Analysis}
Adding up all the contributions considered in the previous sections and
neglecting  $(a_l)^H_{\rm \,SM}$,
the supersymmetric contribution to the muon MDM is given by
\cite{Stockinger:2006zn}
\begin{eqnarray}
(a_\mu)_{\rm \,SUSY} & =  &
\left(1-\frac{4 \alpha_{\rm em}}{\pi}\,\log\frac{M_{\rm SUSY}}{m_\mu} \right)\,
\left[ (a_\mu)^{\chi^\pm} + (a_\mu)^{\chi^0}+(a_\mu)^H \right]
\nonumber \\
& \equiv &
(a_\mu)_{\rm \,SUSY}^{\chi^\pm}
+(a_\mu)_{\rm \,SUSY}^{\chi^0}
+(a_\mu)_{\rm \,SUSY}^{H}\,,
\end{eqnarray}
where the large QED logarithm takes into account 
the renormalization-group (RG) evolution 
of $a_\mu$ from the
SUSY scale down to the muon-mass scale. The logarithmic correction amounts to -7 \% and
-9 \% for $M_{\rm SUSY}=100$ GeV and 1000 GeV, respectively.

{\subsection{ A typical scenario where 1 loop dominates}}

We first consider a typical scenario in which the dominant contributions come from the
one-loop chargino and neutralino diagrams~\cite{Stockinger:2006zn}.  We set
\begin{equation}
|\mu| = |M_2| = 2 |M_1| = M_{\tilde{L}_2} = M_{\tilde{E}_2} = M_S\,.
\label{eq:S1_1}
\end{equation}
The common scale $M_S$ and $\tan\beta$ are varied. 
For CP phases, we first consider the two values for
$\Phi_\mu = 0^\circ~{\rm or}~180^\circ$  while taking vanishing CP phases for the 
gaugino mass and $A$ parameters:
$\Phi_{1,2,3}=\Phi_A=0^\circ$. The remaining relevant parameters are fixed as
\begin{eqnarray}
&& M_{H^\pm}=300~~{\rm GeV}, 
\nonumber \\
&& M_{\tilde{Q}_3} = M_{\tilde{U}_3} = M_{\tilde{D}_3} =
M_{\tilde{L}_3} = M_{\tilde{E}_3} = 0.5 ~~{\rm TeV},
\nonumber \\
&& |A_{t,b,\tau\,\mu}|=1 ~~{\rm TeV},   \ \ |M_3|=1 ~~{\rm TeV}\,.
\label{eq:S1_2}
\end{eqnarray}

\begin{figure}[t]
\hspace{ 0.0cm}
\vspace{-0.5cm}
\centerline{\epsfig{figure=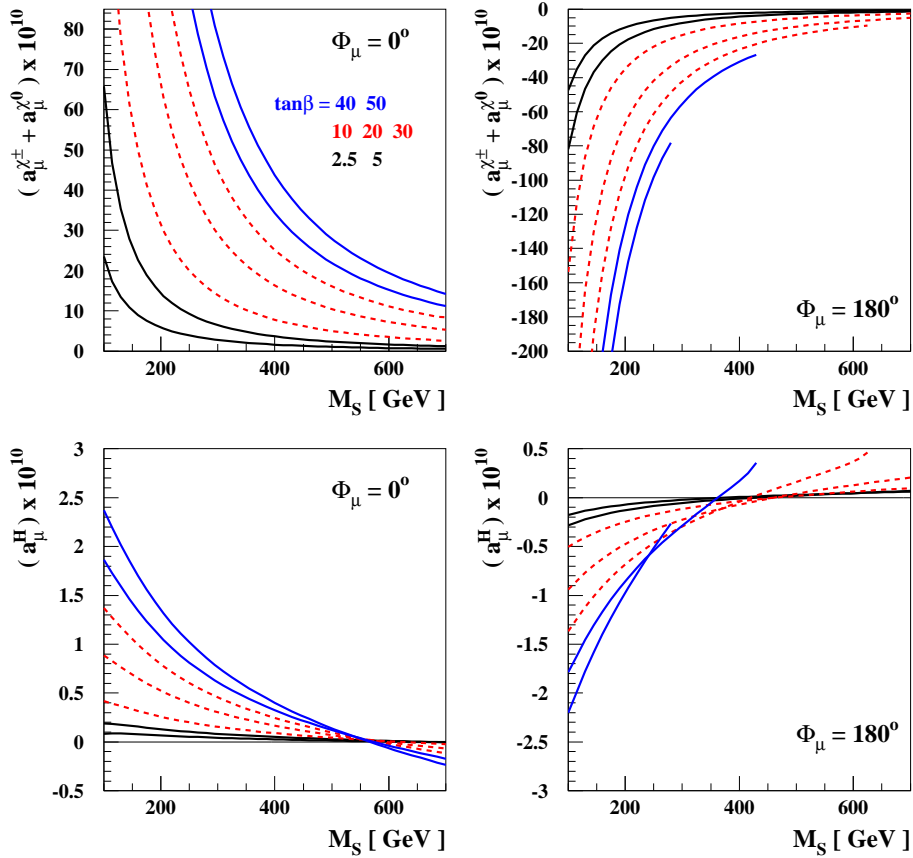,height=13.0cm,width=13.0cm}}
\caption{\it The one- (upper) and two-loop (lower) contributions to $(a_\mu)_{\rm \,SUSY}$
as functions of $M_S$ (\ref{eq:S1_1}) 
for $\tan\beta=2.5\,,5\,,10\,,20\,,30\,,40\,,50$. 
In all frames, the smaller 
value of $\tan\beta$ gives smaller $|a_\mu|$.  
The other parameters are chosen as in Eq.~(\ref{eq:S1_2}).
}
\label{fig:amu.check}
\end{figure}
In Fig.~\ref{fig:amu.check}, we show the one-loop contributions from the 
chargino and
neutralino diagrams (upper) and the two-loop contributions from the Barr-Zee graphs (lower)
to the SUSY muon MDM as functions of $M_S$ for
several values of $\tan\beta$. The left frames are for $\Phi_\mu=0^\circ$ and 
the right ones for $\Phi_\mu=180^\circ$. In both cases,
the two-loop and one-loop contributions have same signs for most regions of $M_S$.
The one-loop and two-loop contributions drops rapidly as $M_S$ increases.
The case of $\Phi_\mu=0^\circ$ gives the correct sign and, for example,
we have $(a_\mu)_{\rm \,SUSY} \times 10^{10} \gsim 10$ at $M_S = 500$ GeV 
when $\tan\beta\gsim 20$.
For $\Phi_\mu=180^\circ$, $(a_\mu)_{\rm \,SUSY}$ is negative
and the considered regions of $M_S$ are not allowed because data 
prefers a positive $a_\mu$.
Note that for $\tan\beta \geq 30$, 
the $\tan\beta$-enhanced threshold corrections can turn
the $b$-quark Yukawa coupling non-perturbative. This happens when 
$M_S$ (or $|\mu|$) is sufficiently large, as for the case at hand.
It is shown by the termination of the curves in the right panels 
of Fig. \ref{fig:amu.check}.
Our results are in good  agreement with existing ones in literature.

\begin{figure}[t]
\hspace{ 0.0cm}
\vspace{-0.5cm}
\centerline{\epsfig{figure=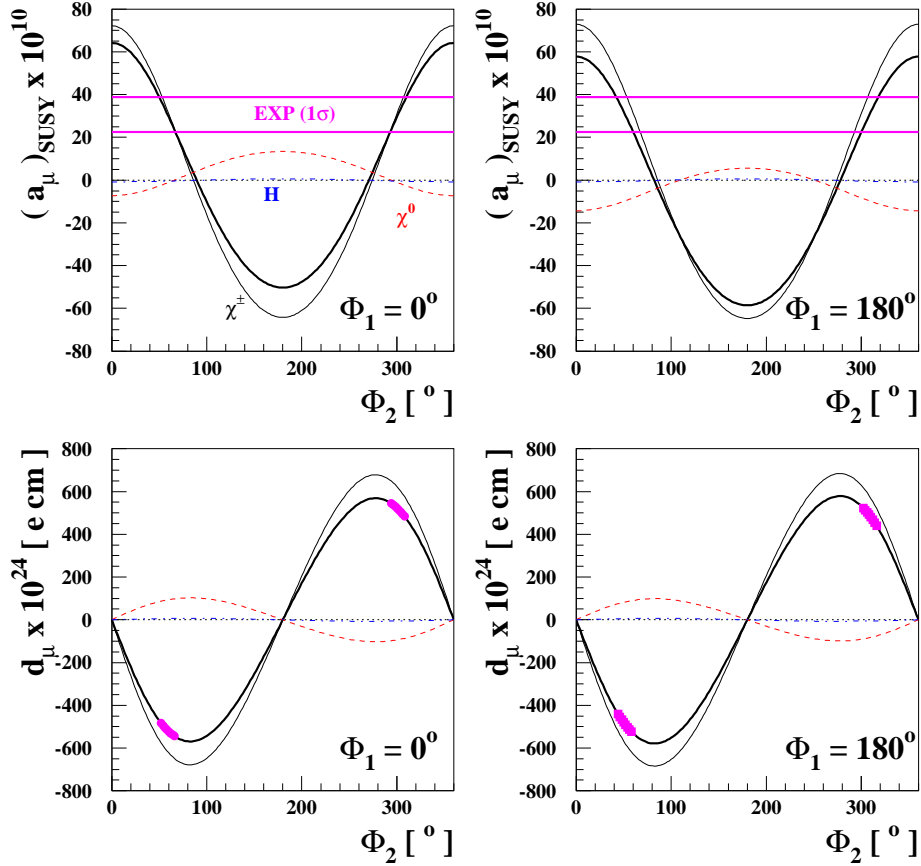,height=13.0cm,width=13.0cm}}
\caption{\it The MDM (upper) and EDM (lower) of the muon
as functions of $\Phi_2$ taking $\Phi_1=0^\circ$ (left) and
$180^\circ$ (right) when $\Phi_\mu=0^\circ$, $M_S=250$ GeV and $\tan\beta=30$.
The thick line is for the total MDM/EDM and the thin solid,
dashed, dash-dotted lines are for the constituent contributions from 
the one-loop chargino, the one-loop neutralino, 
and the two-loop Barr-Zee diagrams, respectively.
}
\label{fig:admu.p1p2}
\end{figure}
Figure~\ref{fig:admu.p1p2} shows the SUSY muon MDM (upper) and EDM (lower)
as functions of $\Phi_2$ taking $\Phi_1=0^\circ$ (left) and
$180^\circ$ (right).
We have taken $\Phi_\mu=0^\circ$, $M_S=250$ GeV,
$\tan\beta=30$, and the other parameters the same as
in Fig.~\ref{fig:amu.check}.
In both the MDM and EDM,
we observe that the dominant contribution is
coming from the one-loop chargino diagrams.
The subleading contribution from the neutralino diagrams is about
5 to 10 times smaller and likely has an opposite sign with respect to the 
dominant chargino contribution. The contributions from the Higgs-mediated 
Barr-Zee diagrams are negligible. Numerically, we have 
$-0.8 < (a_\mu)_{\rm \,SUSY}^H \times 10^{10} < 0.6$ and
$|(d_\mu)^H| \times 10^{24} \lsim 7\,e\,\rm{cm}$.
We clearly see the (shifted) cosine and sine functional forms of 
the MDM and EDM, respectively, as to be anticipated from
the relations given in Eq.~(\ref{eq:oneloop}).
In the upper frames, the horizontal band is the experimental
$1$-$\sigma$-allowed region, $(30.7\pm 8.2)\times 10^{10}$;
 see Eq.~(\ref{eq:damu}). 
In the lower frames the region is overlayed with thick dots
along the thick solid line. 
We note the chosen parameter set is compatible with the experimental data only 
for the non-trivial values of $\Phi_2$ around $60^\circ$ and $300^\circ$,
resulting in large EDM of about $\pm 5 \times 10^{-22}\, e\,$cm,
which can be easily observed
once the projected sensitivity of $10^{-24}$ can be achieved.
%

\begin{figure}[t]
\hspace{ 0.0cm}
\vspace{-0.5cm}
\centerline{\epsfig{figure=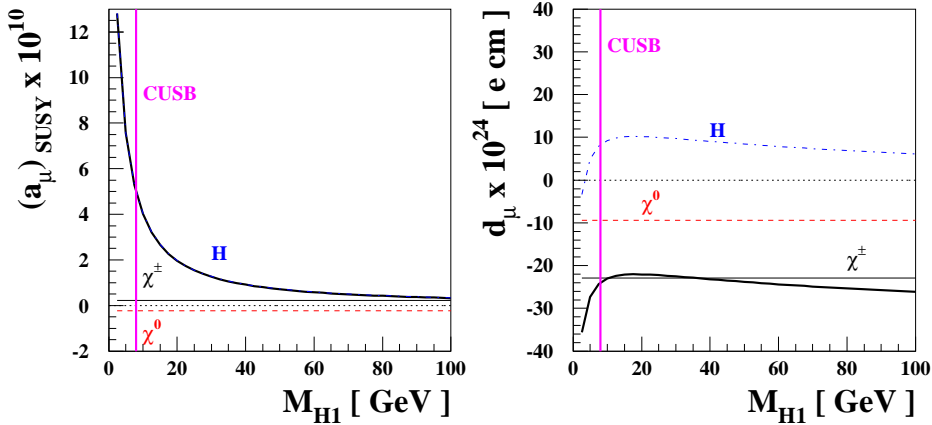,height=13.0cm,width=13.0cm}}
\vspace{-6.5cm}
\caption{\it The muon SUSY MDM (left) and EDM (right)
as functions of $M_{H_1}$ in the CPX scenario with $\tan\beta=10$, see
Eq.~(\ref{eq:CPX}). The lines are the same as in
Fig.~\ref{fig:admu.p1p2}.  
In each frame, the region left to the vertical line
is excluded by data on $\Upsilon(1S)$ decay\protect\cite{upsilon_visible}.
}
\label{fig:admu.CPX} \end{figure}

\subsection{CPX scenario}

Next we consider the CPX scenario\cite{Carena:2000ks}:
\begin{eqnarray}
&& \hspace{-3.0cm}
M_{\tilde{Q}_3} = M_{\tilde{U}_3} = M_{\tilde{D}_3} =
M_{\tilde{L}_3} = M_{\tilde{E}_3} = M_{\rm SUSY}\,,
\nonumber \\
&& \hspace{-3.0cm}
|\mu|=4\,M_{\rm SUSY}\,, \ \
|A_{t,b,\tau}|=2\,M_{\rm SUSY} \,, \ \
|M_3|=1 ~~{\rm TeV}\,.
\label{eq:CPX}
\end{eqnarray}
Taking $A_\mu=A_\tau$,
we fixed  
$\Phi_A=\Phi_3=90^\circ$, $M_{\rm SUSY}=0.5$ TeV,
$|M_2|=2|M_1|=100$ GeV with $\Phi_{1,2}=90^\circ$,
and $M_{\tilde{L}_2} = M_{\tilde{E}_2} = M_{\rm SUSY}$. 
For our analysis,
the most relevant feature of the scenario is that
the combined searches of the four LEP collaborations reported two allowed regions where
the lightest Higgs boson $H_1$ can be very light
for moderate values of $3 \lsim \tan\beta \lsim 10$~\cite{Schael:2006cr}:
\begin{eqnarray}
    && M_{H_1} \lsim 10 \;\; {\rm GeV}\;\;\;\;
      {\rm for}\;\; 3 \lsim \tan\beta \lsim 10,\nonumber\\
    &&30\;\; {\rm GeV} \lsim M_{H_1} \lsim 50\;\; {\rm
    GeV}\;\;\;\; {\rm for} \;\; 3  \lsim \tan\beta \lsim 10\,.
\label{eq:lep}
\end{eqnarray}
\noindent
On the other hand, a lower limit on the lightest 
Higgs boson, $M_{H_1} \gsim 8$ GeV, is available from the bottomonium decay
$\Upsilon(1S)\rightarrow \gamma H_1$~\cite{upsilon_visible}.
Figure~\ref{fig:admu.CPX} shows $(a_\mu)_{\rm \,SUSY}$ and $d_\mu$ 
in the CPX scenario as functions of $M_{H_1}$ taking $\tan\beta=10$. 
When $M_{H_1} \lsim 50$ GeV,
the one-loop contributions to $(a_\mu)_{\rm \,SUSY}$ are negligible
compared to the Higgs-mediated two-loop contributions.
The sign of $(a_\mu)^H_{\rm \,SUSY}$ is plus ($+$)
since it is dominated by the bottom-quark and tau-lepton loops
mediated by $H_1$ which is almost the CP odd state; 
see Eq.~(\ref{eq:ah}). However, it is still difficult
to achieve $(a_\mu)_{\rm \,SUSY} \times 10^{10} \gsim +10$
only with a mostly CP-odd Higgs boson as light as $\sim$ 8 GeV.
For the EDM, the one- and two-loop contributions are comparable and tend to
cancel each other.

\subsection{An extreme scenario} 

Finally, we consider a scenario in which the one-loop neutralino and 
chargino contributions are
suppressed while the two-loop Barr-Zee contributions dominate.
This scenario is characterized by large $\tan\beta$, a light charged 
Higgs boson, very heavy
smuons and muon sneutrinos, and very large $|\mu|$ and 
$|A_{b\,,\tau}|$ parameters. Explicitly, we
have chosen
\begin{eqnarray}
&&\tan\beta=50, \ \ M_{H^\pm}=0.2~~{\rm TeV},
\nonumber \\
&&M_{\tilde{L}_2} = M_{\tilde{E}_2} = 10 ~~{\rm TeV},
\nonumber \\
&&M_{\tilde{Q}_3} = M_{\tilde{U}_3} = M_{\tilde{D}_3} =
M_{\tilde{L}_3} = M_{\tilde{E}_3} = 1 ~~{\rm TeV},
\nonumber \\
&& 
|A_t|=1 ~~{\rm TeV}, \ \
|M_2|=2\,|M_1|=0.1~~{\rm TeV}, \ \ |M_3|=1 ~~{\rm TeV},
\nonumber \\
&&
\Phi_\mu = 0^\circ\,, \ \ \Phi_{A_t}=0^\circ\,, \ \ \Phi_{1,2}=0^\circ\,,
\label{eq:EXT1}
\end{eqnarray}
while varying
\begin{eqnarray}
\Phi_{A_{b\tau}}\,, \ \ \Phi_3\,; \ \
1 < |\mu|/{\rm TeV} < 12\,, \ \
1 < |A_{b\tau}|/{\rm TeV} < 50\,.
\label{eq:EXT2}
\end{eqnarray}
where $|A_{b\tau}|\equiv |A_b| = |A_\tau|$ and we have taken $A_\mu=A_\tau$.
Note $|A_t|$ is fixed in this scenario and the results are almost independent of
$\Phi_{1,2}$.

\begin{figure}[t]
\hspace{ 0.0cm}
\vspace{-0.5cm}
\centerline{\epsfig{figure=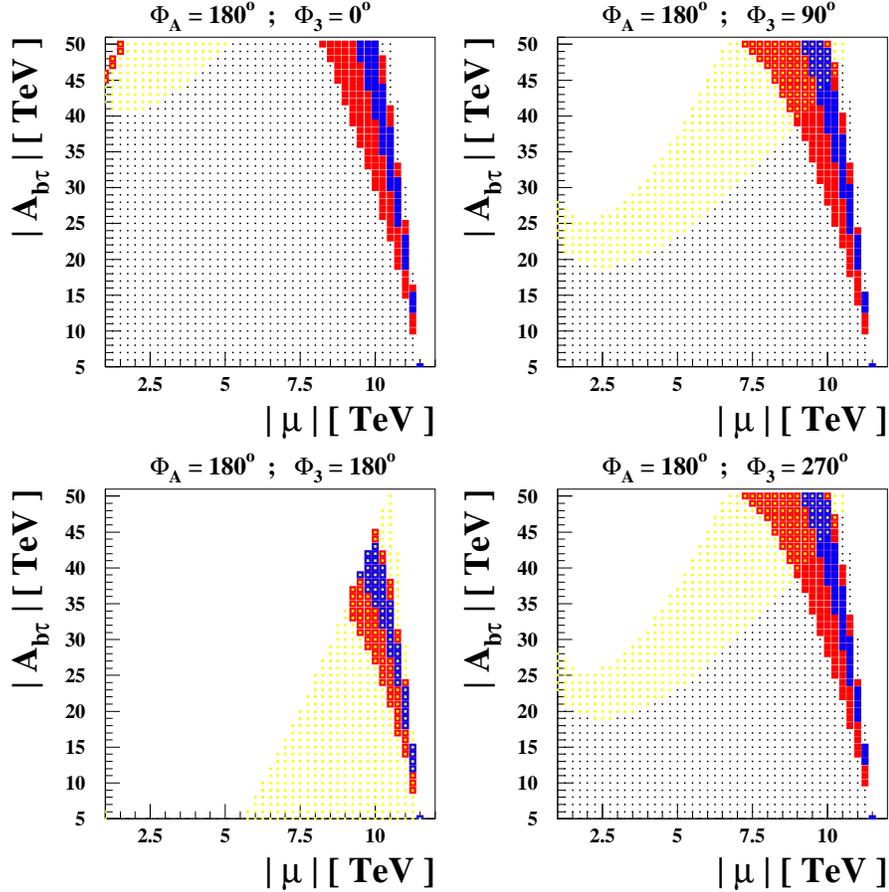,height=13.0cm,width=13.0cm}}
\caption{\it Allowed region at the 1-$\sigma$ (blue) and 2-$\sigma$ (blue+red) level in
the $|A_{b\tau}|$ and $|\mu|$ plane, see Eq.~(\ref{eq:damu}), for $\Phi_3=0^\circ$
(upper-left), $90^\circ$ (upper-right), $180^\circ$ (lower-left), and $270^\circ$
(lower-right). We have taken 
$\Phi_{A_{b\tau}}=180^\circ$. The unshaded regions are not
theoretically allowed and $M_{H_1}<100$ GeV in the over-shaded regions (yellow).
See Eq.~(\ref{eq:EXT1}) for other parameters chosen.
}
\label{fig:abtau.mu.EXT} \end{figure}
Figure~\ref{fig:abtau.mu.EXT} shows the regions where
$(a_\mu)_{\rm \,SUSY} = 30.2 \pm 8.8$ (blue)
and $(a_\mu)_{\rm \,SUSY} = 30.2 \pm 17.6$ (blue$+$red)
in the $|A_{b\tau}|$ and $|\mu|$ plane for several values 
of $\Phi_3$, taking $\Phi_{A_{b\tau}}=180^\circ$.
The unshaded regions are not theoretically allowed and 
we have $M_{H_1}<100$ GeV in the over-shaded regions (yellow), for example,
in the upper-left corner in the upper-left frame. We found that $H_1$ is always lighter
than 100 GeV when $\Phi_3=180^\circ$ (lower-left) because
the resummed threshold corrections modify 
the bottom-quark Yukawa coupling significantly in this case.
We note the region with larger $|A_{b\tau}|$ is more preferred.
\begin{figure}[t]
\hspace{ 0.0cm}
\vspace{-0.5cm}
\centerline{\epsfig{figure=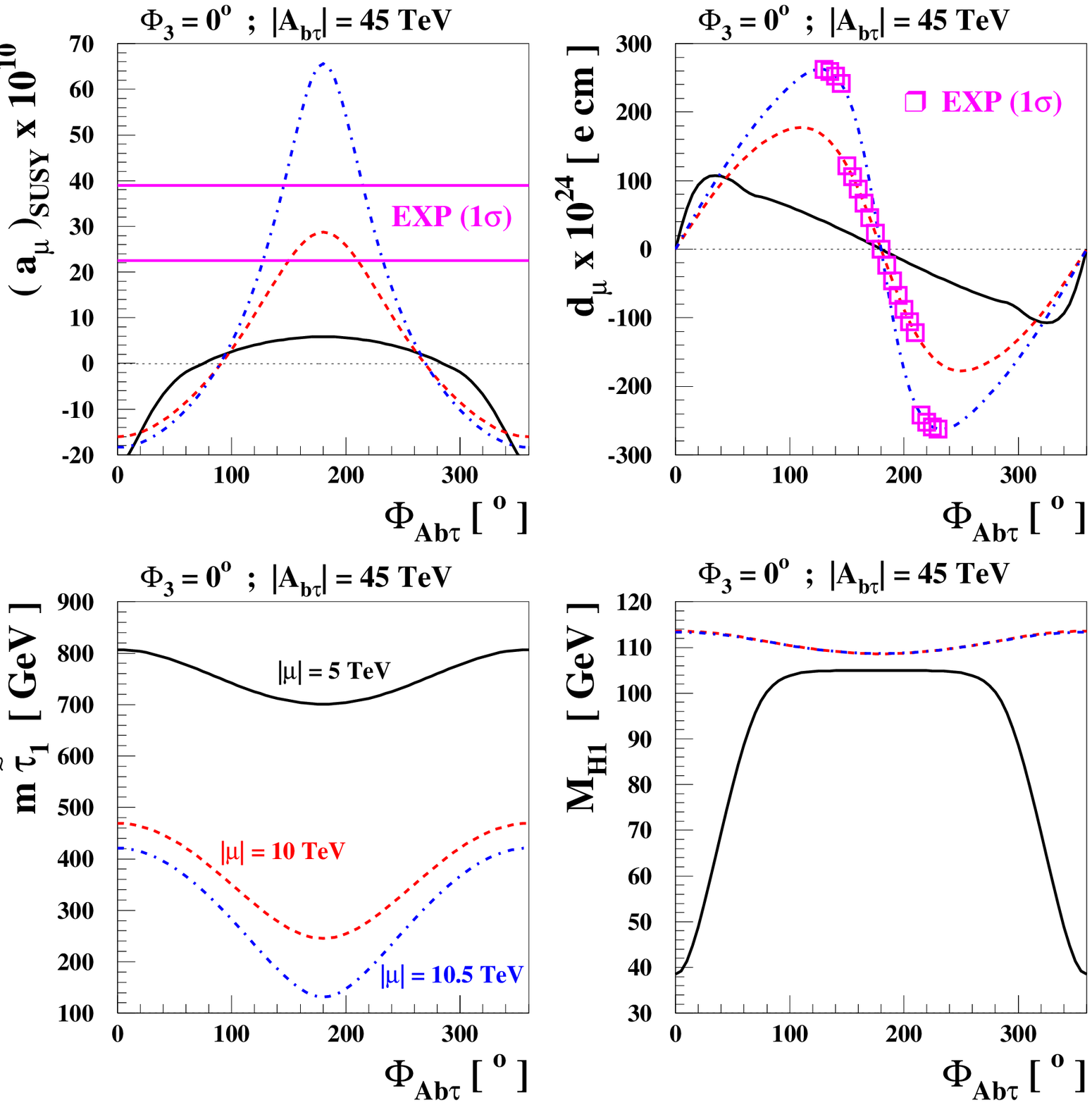,height=13.0cm,width=13.0cm}}
\caption{\it Dependence on $\Phi_{A_{b\tau}}$ of $(a_\mu)_{\rm \,SUSY}$ (upper-left), 
$d_\mu$ (upper-right), and the masses of the lighter stau (lower-left) and $H_1$
(lower-right) for three values of $\mu$: 5 TeV (solid), 10 TeV (dashed), and 10.5 TeV
(dashed-dotted). We have taken 
$\Phi_3=0^\circ$ and $|A_{b\tau}|=45$ TeV.
See Eq.~(\ref{eq:EXT1}) for other parameters chosen.
}
\label{fig:phia.EXT} \end{figure}
Figure~\ref{fig:phia.EXT} shows the dependence on $\Phi_{A_{b\tau}}$ of $(a_\mu)_{\rm
\,SUSY}$, $d_\mu$, and the masses of the lighter stau and the lightest Higgs boson.
In the upper-left frame, the horizontal band is the experimental
$1$-$\sigma$ region of $\Delta a_\mu^{\rm EXP}$.
We found $150^\circ \lsim \Phi_{A_{b\tau}} \lsim 210^\circ$ and 
$\Phi_{A_{b\tau}} \sim 140^\circ\,,220^\circ$ can make
$(a_\mu)_{\rm \,SUSY}$ consistent with the experimental value
for $|\mu|=10$ TeV and 10.5 TeV, respectively. 
In the upper-right frame,
the 1-$\sigma$ region is overlayed with blank boxes along the dashed ($|\mu|=10$ TeV)
and dash-dotted ($|\mu|=10.5$ TeV) lines. 
We have $|d_\mu | \times 10^{24} \lsim 120\,e\,$cm and
$d_\mu \times 10^{24} \sim \pm 250 \,e\,$cm
for $|\mu|=10$ TeV and 10.5 TeV, respectively.
We observe that the larger $|\mu|$ results
in the lighter staus as shown in the lower-left frame.
This leads to a larger $(a_\mu)_{\rm \,SUSY}$ as the dominant contribution 
in this case comes from the 
Higgs-mediated stau Barr-Zee graphs, as will be shown later. 
When  $|\mu|=5$ TeV, $H_1$ becomes lighter than 100
GeV for $\Phi_{A_{b\tau}} \lsim 80^\circ$ and 
$\gsim 280^\circ$, as shown in the lower-right
frame. When $|\mu|\geq 10$ TeV, $M_{H_1} \gsim 108$ GeV.

\begin{figure}[htb]
\hspace{ 0.0cm}
\vspace{-0.5cm}
\centerline{\epsfig{figure=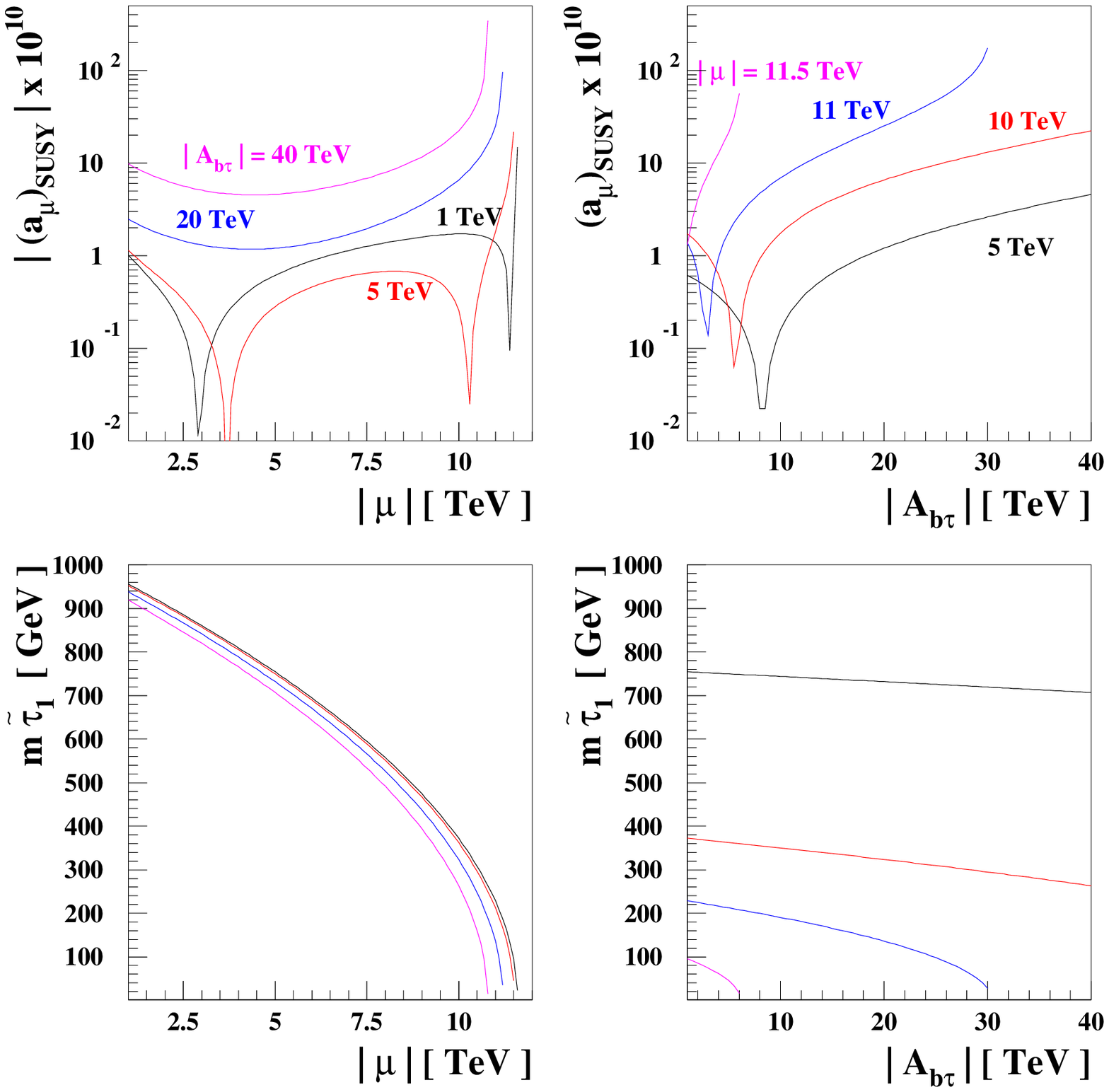,height=13.0cm,width=13.0cm}}
\caption{\it Dependence of $(a_\mu)_{\rm \,SUSY}$ on 
$|\mu|$ (upper-left) and $|A_{b\tau}|$
(upper-right) for several values of $|A_{b\tau}|$ and $|\mu|$, respectively,
taking $\Phi_{A_{b\tau}}=180^\circ$ and $\Phi_3=0^\circ$.
The regions of $|\mu|$ and $|A_{b\tau}|$ are constrained by applying the
experimental limit on the lightest stau mass as shown below. 
The lines are the same as in the upper frames.
See Eq.~(\ref{eq:EXT1}) for other parameters chosen.}
\label{fig:amu.EXT} \end{figure}
Figure~\ref{fig:amu.EXT} shows the dependence of $(a_\mu)_{\rm \,SUSY}$ on
$|\mu|$ (upper-left) and $|A_{b\tau}|$
(upper-right) for several values of $|A_{b\tau}|$ and $|\mu|$, respectively,
taking $\Phi_{A_{b\tau}}=180^\circ$ and $\Phi_3=0^\circ$. In the lower frames we also
show the dependence of the mass of the lighter stau $\tilde\tau_1$.
Again we observe that large $|\mu|$ and $|A_{b\tau}|$ can easily make
$(a_\mu)_{\rm \,SUSY}$ consistent with the current $\Delta a_\mu^{\rm
EXP}$.
\begin{figure}[htb]
\hspace{ 0.0cm}
\vspace{-0.5cm}
\centerline{\epsfig{figure=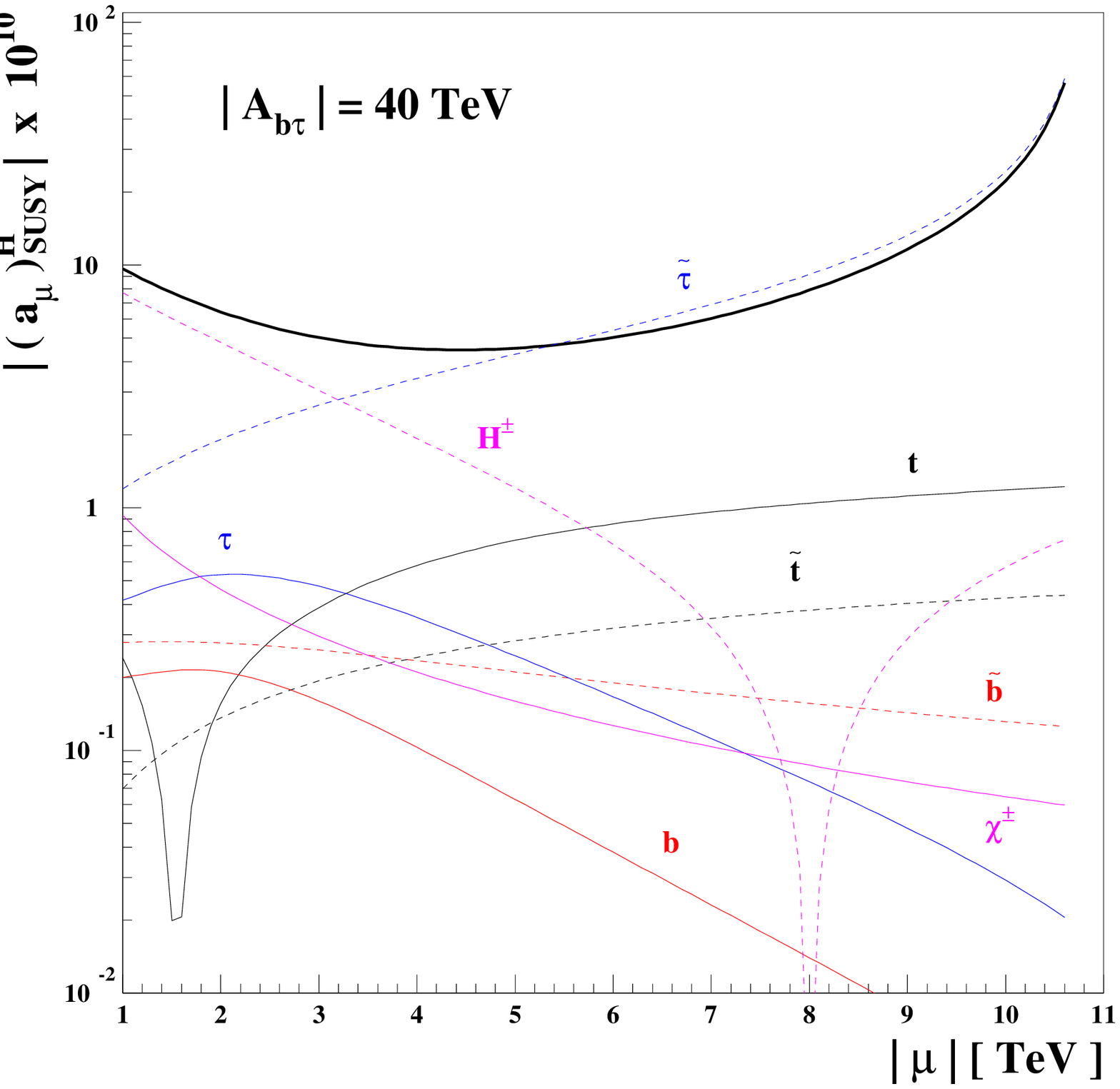,height=13.0cm,width=13.0cm}}
\caption{\it The Barr-Zee contributions to $(a_\mu)_{\rm \,SUSY}$, 
$(a_\mu)_{\rm \,SUSY}^H$, as functions of $|\mu|$
taking $|A_{b\tau}|=40$ TeV, $\Phi_{A_{b\tau}}=180^\circ$ and $\Phi_3=0^\circ$.
Also shown are the constituent contribution from the eight types of diagrams.
See Eq.~(\ref{eq:EXT1}) for other parameters chosen.}
\label{fig:amu4.40.EXT} \end{figure}
Figure~\ref{fig:amu4.40.EXT} shows various constituent as well as the 
total two-loop Barr-Zee contributions to the
muon MDM 
$(a_\mu)_{\rm \,SUSY}^H$, as functions of $|\mu|$
taking $|A_{b\tau}|=40$ TeV, $\Phi_{A_{b\tau}}=180^\circ$ and $\Phi_3=0^\circ$.
The thick line is for the total and the thin lines are for
the constituent eight contributions (see Fig.~\ref{fig:Barr-Zee}).
For smaller $|\mu|$, the dominant contribution comes from the charged Higgs boson loop. 
This is because the $H_i$-$H^+$-$H^-$ couplings have loop-induced enhancement 
from large $|A_b|$ and $|\mu|$~\cite{Pilaftsis:1999qt}.
The possibility of such a significant contribution of the (photon-)Barr-Zee 
diagram with a (closed) charged Higgs boson loop has apparently not been noticed
before. In fact, the particular diagram is typically not included in analyses
at more or less the same level of numerical precision to ours presented here.
As $|\mu|$ grows, however, the contribution from the stau loops is enhanced and thus 
becomes dominant. The contribution from the sbottom loop is not significant because
the resummed threshold corrections suppress the bottom-quark Yukawa coupling
for the chosen parameter set.
\begin{figure}[htb]
\hspace{ 0.0cm}
\vspace{-6.5cm}
\centerline{\epsfig{figure=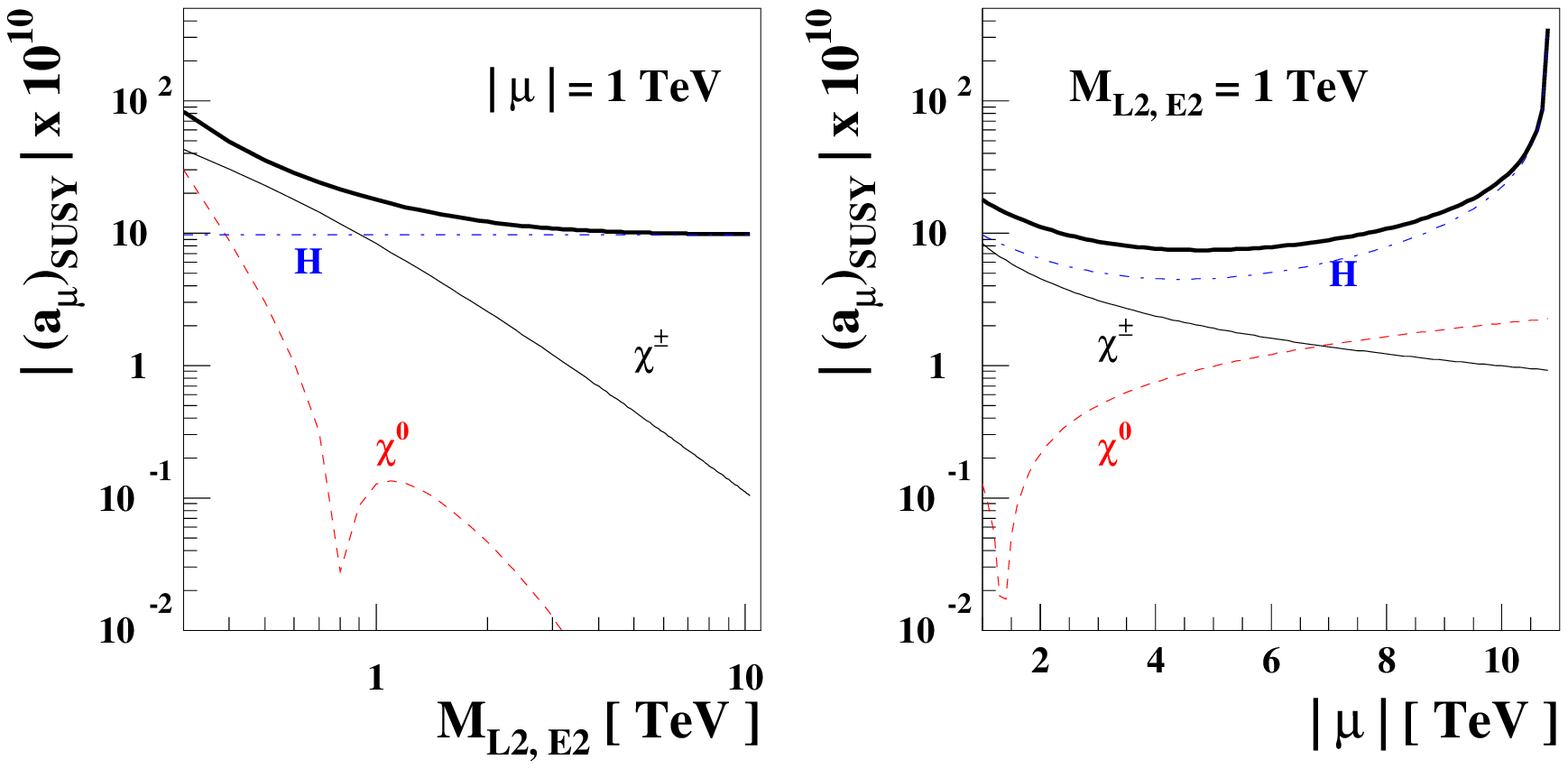,height=13.0cm,width=13.0cm}}
\caption{\it Dependence of $(a_\mu)_{\rm \,SUSY}$ on 
$M_{\tilde{L}_2, \tilde{E}_2}$ (left)
and $|\mu|$ (right) for
fixed values of $|\mu|$ and $M_{\tilde{L}_2, \tilde{E}_2}$. 
The thick line is for the total and the thin solid (black), 
dashed (red), and dash-dotted (blue) lines are for
the constituent contributions from
the one-loop chargino, one-loop neutralino,  and
Higgs-mediated two-loop Barr-Zee diagrams.
See Eq.~(\ref{eq:EXT1}) for other parameters chosen.}
\label{fig:amu.LE.EXT} \end{figure}
The  scenario likely seems too contrived to some readers. However, 
we are presenting it here mainly to illustrate the significant roles of the 
various two-loop Barr-Zee contributions in some region of the parameter space. 
While the dominance of the latter group reduces as one moves away from the
extreme corner of the parameter space, its significance maintains over a
substantial region. For instance, we show the dependence of $(a_\mu)_{\rm \,SUSY}$ on
$M_{\tilde{L}_2, \tilde{E}_2}$ and $|\mu|$ in Fig.~\ref{fig:amu.LE.EXT},
{\it i.e.} the effect of bringing back the smuon mass from the heavy limit. 
In particular, the Higgs-mediated two-loop Barr-Zee contributions are shown to be 
actually dominate over the one-loop contributions even when the smuon mass
parameters
get down to as low as 1 TeV (left), for the full range of $|\mu|$ values (right).

Before closing this section, we comment on the relation between 
the muon and electron EDMs. The most important one-loop contribution $(d_l)^\chi$
and that from the two-loop $(d_l)^H$ have somewhat different features. The
Barr-Zee diagrams for the muon and the electron are identical except for
the muon and the electron lines themselves. Hence, we have the robust
relation
\begin{equation}
(d_e)^H = (m_e/m_\mu)\,(d_\mu)^H\,.
\end{equation}
For the case of $(d_l)^\chi$, however, it is sensitive to the flavor
dependence of the soft SUSY breaking terms. In most of the models on the 
origin of the soft SUSY breaking terms available in the literature, those 
of the first two generations are more or less the same. Explicitly,
\[
M_{\tilde{L}_1} \sim M_{\tilde{L}_2}\,,  \ \
M_{\tilde{E}_1} \sim M_{\tilde{E}_2}\,,  \ \
A_e \sim A_\mu\,. 
\]
That does give us a relation
\begin{equation}
(d_e)^\chi \sim (m_e/m_\mu)\,(d_\mu)^\chi \,,
\end{equation}
which is particularly sensitive to the CP 
phases of the $A_e$ and $A_\mu$ parameters.
%
To summarize, one may consider special cases  with 
($i$) similar or universal soft terms (for electron and muon)
or ($ii$) electron EDM dominated by the two-loop Barr-Zee graphs,
$d_e \sim (d_e)^H$, possibly due to 
much heavier selectron masses.
Depending on situations, $d_\mu\sim (d_\mu)^\chi + (d_\mu)^H$ or $(d_\mu)^H$ 
is very strongly constrained by the Thallium EDM as
\begin{equation}
(i)\, \ |d_\mu| < 3 \times 10^{-25}\,e\,{\rm cm} \ \ ; \ \
(ii)\, \ |(d_\mu)^H| < 3 \times 10^{-25}\,e\,{\rm cm}\,,
\end{equation}
where we used $m_e/m_\mu=1/213$, $d_{\rm Tl} \simeq -585\cdot d_e$
and $|d_{\rm Tl}| < 9\times 10^{-25}\,e\,$cm~\cite{Regan:2002ta}. Note 
that these limits are model dependent and the muon EDM should be measured 
independently of the Thallium EDM.

\section{Conclusions}
We have studied in detail various supersymmetric contributions
to muon MDM and EDM, including one-loop chargino and neutralino diagrams
and dominant two-loop Barr-Zee diagrams. In general, the one-loop
contributions dominate over the two-loop contributions, however there
are interesting regions of the model parameter space where the two-loop 
Barr-Zee diagrams are the major source of contributions to muon MDM and/or EDM.  
The model parameter space is huge. It is not feasible for us to present
and discuss here the numerical results for more than a few cases of interest.
We try to pick cases that can illustrate the essential and interesting 
features, and leave it mostly to the readers to project onto the parameter
space regions in between, or considered otherwise to be of special interest.
We illustrate numerically 3 scenarios under various choices of soft parameters: 
(i) the one when the one-loop contributions dominate,
(ii) the CPX in which the Barr-Zee dominates but the
overall sizes of MDM and EDM are small, and (iii) a more exotic one in 
which the Barr-Zee dominates and the overall sizes of MDM and EDM are large.
We have also shown interesting relations between the MDM and EDM.
For the case when one-loop contributions dominate the MDM and EDM are
just, respectively, the (shifted) cosines and sines of the phase of 
the parameters involved. Existing experimentally preferred range of MDM
already predicts an interesting range of EDM, which can be further
tested in the future muon EDM experiments.  
For the case the MDM is dominated by two-loop Barr-Zee contributions the MDM 
and EDM can be connected by the relation in Eq.(\ref{bbb}) --- a result of the
more complicated Higgs sector phase structure.

The CPX scenario may still allow a light Higgs boson after taking into
account all the existing search limits. Potentially, the light Higgs boson could give
large enhancement to muon MDM.  
However, after imposing the lower limit on $M_{H_1}$ the resulting MDM is always less
than $5\times 10^{-10}$, 
which is smaller than the experimentally favored value. 

The last scenario associated with large $\tan\beta$, large $|\mu|$,
heavy smuons and muon sneutrinos, and large $|A_{b\tau}|$ but light 
charged Higgs boson and stau is rather interesting.  It suppresses
the one-loop contributions but the two-loop contributions are large enough
to explain the muon MDM data, which are dominated by charged Higgs boson
and stau at smaller and larger $|\mu|$, respectively.  For large enough $|\mu|$
the MDM data can be accommodated easily. The particular interesting role of the
charged Higgs boson escaped earlier studies. We have also illustrated that
major features of the scenario persist over a region of the parameter space
with milder conditions --- in particular, more `regular' smuon masses.

In addition, we offer the following comments.
\begin{enumerate}
\item
We have included the threshold corrections to Yukawa couplings. 
In particular, the bottom-quark Yukawa can receive large corrections at
large $\tan\beta$ with large $|M_3|$ and $|\mu|$.  
For $\tan\beta \gsim 30$ the threshold corrections can make
the $b$-quark Yukawa coupling turn non-perturbative when $M_S$ 
(or $|\mu|$) is sufficiently large.

\item The one-loop contributions to MDM and EDM can vary as 
shifted cosines and sines of the phase of the parameters.

\item As shown in Fig. \ref{fig:admu.p1p2}, the prediction for EDM 
is of order $500 \times 10^{-24}\, e\;{\rm cm}$ within the allowed
range of MDM. It is about $2-3$ orders of magnitude below the current limit,
but will be within reach of future muon EDM 
experiments \cite{Semertzidis:1999kv}.

\item The CPX scenario may still allow a $H_1$ as light as a few to tens of GeVs.
It could be searched in the subsequent decay of the $H_2 \to H_1 H_1$, where
$H_2$ is the SM-like Higgs boson.  
The contribution of $H_1$ to the muon EDM has a right sign but 
it may not be large enough to accommodate
$\Delta a_\mu^{\rm EXP}$ after
taking account of the constraint from the bottomonium decay
$\Upsilon(1S)\rightarrow \gamma H_1$.

\item The last scenario that we studied is characterized by a very light
stau, a light bino and wino, a light $M_{H_1}$, and a 
light charged Higgs boson.  The 
predicted EDM is from 0 to $100 \times 10^{-24}\; e\, {\rm cm}$.
Experimental searches for this scenario at the LHC will be 
a lot of tau leptons in the final state because of lightness of stau.

\item The parameter space compatible with the $\Delta a_\mu^{\rm EXP}$ value required
is generally extended by allowing CP phases. 
For example, in Fig.~\ref{fig:admu.p1p2}: 
the CP-conserving cases ($\Phi_2=0^\circ$ and $180^\circ$) are excluded.
\end{enumerate}

\vspace{-0.2cm}
\subsection*{Acknowledgements} 
\vspace{-0.3cm}
\noindent
We thank Dominik St\"ockinger and Apostolos Pilaftsis
for helpful discussions.  
We also thank Joaquim Prades for valuable information on the hadronic
light-by-light contributions.
The work was partially 
supported by the NSC of Taiwan (96-2112-M-008-007-MY3), the Boost
Project of NTHU, and the WCU program through the KOSEF funded by
the MEST (R31-2008-000-10057-0).

%

\def\theequation{\Alph{section}.\arabic{equation}}
\begin{appendix}
\setcounter{equation}{0}
\section{{\tt CPsuperH} Interface}
\label{sec:cpsuperh}
\begin{itemize}
\item \underline{Output}: For output, part of auxiliary array
{\tt RAUX\_H} is used.
\begin{itemize}
\item{The muon EDM in units of $cm$}:
\begin{equation}
{\tt RAUX\_H(360)} = d_\mu/e = (d_\mu/e)^{\tilde{\chi}^\pm}
+(d_\mu/e)^{\tilde{\chi}^0}+(d_\mu/e)^{\tilde{g}}+(d_\mu/e)^{H} \,, 
\end{equation}
where the sub-contributions are
\begin{eqnarray}
{\tt RAUX\_H(361)} &=& (d_\mu/e)^{\tilde{\chi}^\pm}\,, \ \ \
{\tt RAUX\_H(362)}  =  (d_\mu/e)^{\tilde{\chi}^0}\, \nonumber \\
{\tt RAUX\_H(363)} &=& (d_\mu/e)^{\tilde{g}}\,, \ \ \ \ \
{\tt RAUX\_H(364)}  =  (d_\mu/e)^{H}\,. \nonumber
\end{eqnarray}
\item{The muon MDM}:
\begin{equation}
{\tt RAUX\_H(380)} = (a_\mu)_{\rm \,SUSY} = (a_\mu)_{\rm \,SUSY}^{\tilde{\chi}^\pm}
+(a_\mu)_{\rm \,SUSY}^{\tilde{\chi}^0}+(a_\mu)_{\rm \,SUSY}^{\tilde{g}}
+(a_\mu)_{\rm \,SUSY}^{H} \,,
\end{equation}
where the sub-contributions are
\begin{eqnarray}
{\tt RAUX\_H(381)} &=& (a_\mu)_{\rm \,SUSY}^{\tilde{\chi}^\pm}\,, \ \ \
{\tt RAUX\_H(382)}  =  (a_\mu)_{\rm \,SUSY}^{\tilde{\chi}^0}\, \nonumber \\
{\tt RAUX\_H(383)} &=& (a_\mu)_{\rm \,SUSY}^{\tilde{g}}\,, \ \ \ 
{\tt RAUX\_H(384)}  =  (a_\mu)_{\rm \,SUSY}^{H}\,. \nonumber
\end{eqnarray}
\end{itemize}

\item {\tt IFLAG\_H(19)}=1 
is used to print out EDM/MDM of the muon. Using {\tt run} shell-script file distributed,
the sample out obtained is \\ \\
{\tt
~---------------------------------------------------------\\
$~$The~Electric~EDM~of~muon~in~cm:\\
~---------------------------------------------------------\\
$~~~~~$d\^{}E\_mu/e[Total]:~~0.1288E-23\\
$~~$d\^{}E\_mu/e[C,N,Gl,H]:~~0.0000E+00~~-.6035E-24~~0.0000E+00~~0.1891E-23\\
~---------------------------------------------------------\\
$~$The~SUSY~MDM~of~muon:\\
~---------------------------------------------------------\\
$~~~~~~~~~$a\_mu[Total]:~~0.1805E-09\\
$~~~~~~$a\_mu[C,N,Gl,H]:~~0.1396E-09~~0.2871E-10~~0.0000E+00~~0.1218E-10\\
~---------------------------------------------------------\\
} 
\end{itemize}
\end{appendix}


%
%


\begin{thebibliography}{99}

\bibitem{ref:exp}
  R.~M.~Carey {\it et al.},
  ``New measurement of the anomalous magnetic moment of the positive muon,''
  Phys.\ Rev.\ Lett.\  {\bf 82} (1999) 1632;
  H.~N.~Brown {\it et al.}  [Muon (g-2) Collaboration],
  ``Improved measurement of the positive muon anomalous magnetic moment,''
  Phys.\ Rev.\  D {\bf 62} (2000) 091101
  [arXiv:hep-ex/0009029];
  H.~N.~Brown {\it et al.}  [Muon g-2 Collaboration],
  ``Precise measurement of the positive muon anomalous magnetic moment,''
  Phys.\ Rev.\ Lett.\  {\bf 86} (2001) 2227
  [arXiv:hep-ex/0102017];
  G.~W.~Bennett {\it et al.}  [Muon g-2 Collaboration],
  ``Measurement of the Positive Muon Anomalous Magnetic Moment to 0.7 ppm,''
  Phys.\ Rev.\ Lett.\  {\bf 89} (2002) 101804
  [Erratum-ibid.\  {\bf 89} (2002) 129903]
  [arXiv:hep-ex/0208001];
  G.~W.~Bennett {\it et al.}  [Muon g-2 Collaboration],
  ``Measurement of the negative muon anomalous magnetic moment to 0.7-ppm,''
  Phys.\ Rev.\ Lett.\  {\bf 92} (2004) 161802
  [arXiv:hep-ex/0401008];
  G.~W.~Bennett {\it et al.}  [Muon G-2 Collaboration],
  ``Final report of the muon E821 anomalous magnetic moment measurement at
  BNL,''
  Phys.\ Rev.\  D {\bf 73} (2006) 072003
  [arXiv:hep-ex/0602035].

\bibitem{SM:REVIEW}
  M.~Passera, W.~J.~Marciano and A.~Sirlin,
  ``The muon g-2 discrepancy: errors or new physics?,''
  arXiv:0809.4062 [hep-ph];
  F.~Jegerlehner and A.~Nyffeler,
  ``The Muon g-2,''
  arXiv:0902.3360 [hep-ph];
  J.~P.~Miller, E.~de Rafael and B.~L.~Roberts,
  ``Muon g-2: Review of Theory and Experiment,''
  Rept.\ Prog.\ Phys.\  {\bf 70} (2007) 795
  [arXiv:hep-ph/0703049];
  F.~Jegerlehner,
  ``Essentials of the Muon g-2,''
  Acta Phys.\ Polon.\  B {\bf 38} (2007) 3021
  [arXiv:hep-ph/0703125].

\bibitem{SM:QED1}
  T.~Kinoshita and M.~Nio,
  ``Improved alpha**4 term of the muon anomalous magnetic moment,''
  Phys.\ Rev.\  D {\bf 70} (2004) 113001
  [arXiv:hep-ph/0402206];
  T.~Kinoshita and M.~Nio,
  ``Improved alpha**4 term of the electron anomalous magnetic moment,''
  Phys.\ Rev.\  D {\bf 73} (2006) 013003
  [arXiv:hep-ph/0507249];
  T.~Kinoshita and M.~Nio,
  ``The tenth-order QED contribution to the lepton g-2: Evaluation of  dominant
  alpha**5 terms of muon g-2,''
  Phys.\ Rev.\  D {\bf 73} (2006) 053007
  [arXiv:hep-ph/0512330];
  T.~Aoyama, M.~Hayakawa, T.~Kinoshita and M.~Nio,
  ``Revised value of the eighth-order electron g-2,''
  Phys.\ Rev.\ Lett.\  {\bf 99} (2007) 110406
  [arXiv:0706.3496 [hep-ph]];
  T.~Aoyama, M.~Hayakawa, T.~Kinoshita and M.~Nio,
  ``Revised value of the eighth-order QED contribution to the anomalous
  magnetic moment of the electron,''
  Phys.\ Rev.\  D {\bf 77} (2008) 053012
  [arXiv:0712.2607 [hep-ph]];

\bibitem{SM:QED2}
  S.~Laporta and E.~Remiddi,
  ``The Analytical Value Of The Electron Light-Light Graphs Contribution To The
  Muon (G-2) In QED,''
  Phys.\ Lett.\  B {\bf 301} (1993) 440;
  S.~Laporta and E.~Remiddi,
  ``The analytical value of the electron (g-2) at order alpha~3 in QED,''
  Phys.\ Lett.\  B {\bf 379} (1996) 283
  [arXiv:hep-ph/9602417].

\bibitem{SM:QED3}
  M.~Passera,
  ``Precise mass-dependent QED contributions to leptonic g-2 at order alpha$^2$
  and alpha$^3$,''
  Phys.\ Rev.\  D {\bf 75} (2007) 013002
  [arXiv:hep-ph/0606174];
  A.~L.~Kataev,
  ``Reconsidered estimates of the 10th order QED contributions to the muon
  anomaly,''
  Phys.\ Rev.\  D {\bf 74} (2006) 073011
  [arXiv:hep-ph/0608120].

\bibitem{SM:EW}
  A.~Czarnecki, B.~Krause and W.~J.~Marciano,
  ``Electroweak Fermion loop contributions to the muon anomalous magnetic
  moment,''
  Phys.\ Rev.\  D {\bf 52} (1995) 2619
  [arXiv:hep-ph/9506256];
  A.~Czarnecki, B.~Krause and W.~J.~Marciano,
  ``Electroweak corrections to the muon anomalous magnetic moment,''
  Phys.\ Rev.\ Lett.\  {\bf 76} (1996) 3267
  [arXiv:hep-ph/9512369];
  A.~Czarnecki, W.~J.~Marciano and A.~Vainshtein,
  ``Refinements in electroweak contributions to the muon anomalous magnetic
  moment,''
  Phys.\ Rev.\  D {\bf 67} (2003) 073006
  [Erratum-ibid.\  D {\bf 73} (2006) 119901]
  [arXiv:hep-ph/0212229].

\bibitem{SM:HLO1}
  M.~Davier,
  ``The hadronic contribution to (g-2)(mu),''
  Nucl.\ Phys.\ Proc.\ Suppl.\  {\bf 169} (2007) 288
  [arXiv:hep-ph/0701163];
  S.~Eidelman,
  ``Status Of (G(Mu) - 2)/2 In Standard Model,''
  Acta Phys.\ Polon.\  B {\bf 38} (2007) 3015.

\bibitem{SM:HLO2}
  K.~Hagiwara, A.~D.~Martin, D.~Nomura and T.~Teubner,
  ``Improved predictions for g-2 of the muon and $\alpha_{\rm QED}(M_Z^2)$,''
  Phys.\ Lett.\  B {\bf 649} (2007) 173
  [arXiv:hep-ph/0611102].

\bibitem{SM:HLO3}
  F.~Jegerlehner,
  ``Precision measurements of sigma(hadronic) for alpha(eff)(E) at ILC energies
  and (g-2)(mu),''
  Nucl.\ Phys.\ Proc.\ Suppl.\  {\bf 162} (2006) 22
  [arXiv:hep-ph/0608329].

\bibitem{SM:HLO4}
  J.~F.~de Troconiz and F.~J.~Yndurain,
  ``The hadronic contributions to the anomalous magnetic moment of the  muon,''
  Phys.\ Rev.\  D {\bf 71} (2005) 073008
  [arXiv:hep-ph/0402285].

\bibitem{SM:HLO5}
  F.~Jegerlehner,
  ``Muon g - 2 update,''
  Nucl.\ Phys.\ Proc.\ Suppl.\  {\bf 181-182} (2008) 26.

\bibitem{SM:HVP}
  B.~Krause,
  ``Higher-order hadronic contributions to the anomalous magnetic moment of
  leptons,''
  Phys.\ Lett.\  B {\bf 390} (1997) 392
  [arXiv:hep-ph/9607259].

\bibitem{SM:HLBL1}
  M.~Knecht and A.~Nyffeler,
  ``Hadronic light-by-light corrections to the muon g-2: The pion-pole
  contribution,''
  Phys.\ Rev.\  D {\bf 65} (2002) 073034
  [arXiv:hep-ph/0111058];
  M.~Knecht, A.~Nyffeler, M.~Perrottet and E.~De Rafael,
  ``Hadronic light-by-light scattering contribution to the muon g-2: An
  effective field theory approach,''
  Phys.\ Rev.\ Lett.\  {\bf 88} (2002) 071802
  [arXiv:hep-ph/0111059].

\bibitem{SM:HLBL2}
  K.~Melnikov and A.~Vainshtein,
  ``Hadronic light-by-light scattering contribution to the muon anomalous
  magnetic moment revisited,''
  Phys.\ Rev.\  D {\bf 70} (2004) 113006
  [arXiv:hep-ph/0312226].

\bibitem{SM:HLBL3}
  J.~Bijnens and J.~Prades,
  ``The hadronic light-by-light contribution to the muon anomalous magnetic
  moment: Where do we stand?,''
  Mod.\ Phys.\ Lett.\  A {\bf 22} (2007) 767
  [arXiv:hep-ph/0702170].

\bibitem{SM:HLBL3.5}
  A.~Nyffeler,
  ``Hadronic light-by-light scattering in the muon g-2: a new short-distance
  constraint on pion-exchange,''
  arXiv:0901.1172 [hep-ph].

\bibitem{SM:HLBL4}
  J.~Prades, E.~de Rafael and A.~Vainshtein,
  ``Hadronic Light-by-Light Scattering Contribution to the Muon Anomalous
  Magnetic Moment,''
  arXiv:0901.0306 [hep-ph].

\bibitem{SM:HLBL5}
  D.~K.~Hong and D.~Kim,
  ``Pseudo scalar contributions to light-by-light correction of muon g-2 in
  AdS/QCD,''
  arXiv:0904.4042 [hep-ph].

\bibitem{one_loop_MDM:earlier}
  J.~A.~Grifols and A.~Mendez,
  ``Constraints On Supersymmetric Particle Masses From (G-2) Mu,''
  Phys.\ Rev.\  D {\bf 26} (1982) 1809;
  J.~R.~Ellis, J.~S.~Hagelin and D.~V.~Nanopoulos,
  ``Spin 0 Leptons And The Anomalous Magnetic Moment Of The Muon,''
  Phys.\ Lett.\  B {\bf 116} (1982) 283;
  R.~Barbieri and L.~Maiani,
  ``The Muon Anomalous Magnetic Moment In Broken Supersymmetric Theories,''
  Phys.\ Lett.\  B {\bf 117} (1982) 203.

\bibitem{one_loop_MDM:later1}
  D.~A.~Kosower, L.~M.~Krauss and N.~Sakai,
  ``Low-Energy Supergravity And The Anomalous Magnetic Moment Of The Muon,''
  Phys.\ Lett.\  B {\bf 133} (1983) 305;
  T.~C.~Yuan, R.~L.~Arnowitt, A.~H.~Chamseddine and P.~Nath,
  ``Supersymmetric Electroweak Effects On G-2 (Mu),''
  Z.\ Phys.\  C {\bf 26} (1984) 407;
  J.~C.~Romao, A.~Barroso, M.~C.~Bento and G.~C.~Branco,
  Nucl.\ Phys.\  B {\bf 250} (1985) 295;
  I.~Vendramin,
  ``Constraints On Supersymmetric Parameters From Muon Magnetic Moment,''
  Nuovo Cim.\  A {\bf 101} (1989) 731.

\bibitem{one_loop_MDM:later2}
  S.~A.~Abel, W.~N.~Cottingham and I.~B.~Whittingham,
  ``The Muon magnetic moment in flipped SU(5),''
  Phys.\ Lett.\  B {\bf 259} (1991) 307;
  J.~L.~Lopez, D.~V.~Nanopoulos and X.~Wang,
  ``Large (G-2)-Mu In SU(5) X U(1) Supergravity Models,''
  Phys.\ Rev.\  D {\bf 49} (1994) 366
  [arXiv:hep-ph/9308336];
  U.~Chattopadhyay and P.~Nath,
  ``Probing supergravity grand unification in the Brookhaven g-2 experiment,''
  Phys.\ Rev.\  D {\bf 53} (1996) 1648
  [arXiv:hep-ph/9507386].

\bibitem{one_loop_MDM:recent1}
  T.~Moroi,
  ``The Muon Anomalous Magnetic Dipole Moment in the Minimal Supersymmetric
  Standard Model,''
  Phys.\ Rev.\  D {\bf 53} (1996) 6565
  [Erratum-ibid.\  D {\bf 56} (1997) 4424]
  [arXiv:hep-ph/9512396].

\bibitem{one_loop_MDM:recent2}
  G.~C.~Cho, K.~Hagiwara and M.~Hayakawa,
  ``Muon g-2 and precision electroweak physics in the MSSM,''
  Phys.\ Lett.\  B {\bf 478} (2000) 231
  [arXiv:hep-ph/0001229].

\bibitem{one_loop_MDM:recent2.5}
  S.~P.~Martin and J.~D.~Wells,
  ``Muon anomalous magnetic dipole moment in supersymmetric theories,''
  Phys.\ Rev.\  D {\bf 64} (2001) 035003
  [arXiv:hep-ph/0103067].

\bibitem{Heinemeyer:2003dq}
  S.~Heinemeyer, D.~Stockinger and G.~Weiglein,
  ``Two-loop SUSY corrections to the anomalous magnetic moment of the muon,''
  Nucl.\ Phys.\  B {\bf 690} (2004) 62
  [arXiv:hep-ph/0312264].

\bibitem{Heinemeyer:2004yq}
  S.~Heinemeyer, D.~Stockinger and G.~Weiglein,
  ``Electroweak and supersymmetric two-loop corrections to (g-2)(mu),''
  Nucl.\ Phys.\  B {\bf 699} (2004) 103
  [arXiv:hep-ph/0405255].

\bibitem{Degrassi:1998es}
  G.~Degrassi and G.~F.~Giudice,
  ``QED logarithms in the electroweak corrections to the muon anomalous
  magnetic moment,''
  Phys.\ Rev.\  D {\bf 58} (1998) 053007
  [arXiv:hep-ph/9803384].

\bibitem{Feng:2006ei}
  T.~F.~Feng, X.~Q.~Li, L.~Lin, J.~Maalampi and H.~S.~Song,
  ``The two-loop supersymmetric corrections to lepton anomalous magnetic  and
  electric dipole moments,''
  Phys.\ Rev.\  D {\bf 73} (2006) 116001
  [arXiv:hep-ph/0604171].

\bibitem{Marchetti:2008hw}
  S.~Marchetti, S.~Mertens, U.~Nierste and D.~Stockinger,
  ``Tan(beta)-enhanced supersymmetric corrections to the anomalous magnetic
  moment of the muon,''
  arXiv:0808.1530 [hep-ph].

\bibitem{Feng:2008cn}
  T.~F.~Feng, L.~Sun and X.~Y.~Yang,
  ``Electroweak and supersymmetric two-loop corrections to lepton anomalous
  magnetic and electric dipole moments,''
  Nucl.\ Phys.\  B {\bf 800} (2008) 221
  [arXiv:0805.1122 [hep-ph]].

\bibitem{Barr-Zee_MDM}
  D.~Chang, W.~F.~Chang, C.~H.~Chou and W.~Y.~Keung,
  ``Large two-loop contributions to g-2 from a generic pseudoscalar boson,''
  Phys.\ Rev.\  D {\bf 63} (2001) 091301
  [arXiv:hep-ph/0009292];

  K.~m.~Cheung, C.~H.~Chou and O.~C.~W.~Kong,
  ``Muon anomalous magnetic moment, two-Higgs-doublet model, and
  supersymmetry,''
  Phys.\ Rev.\  D {\bf 64} (2001) 111301
  [arXiv:hep-ph/0103183];
  A.~Arhrib and S.~Baek,
  ``Two-loop Barr-Zee type contributions to (g-2)(mu) in the MSSM,''
  Phys.\ Rev.\  D {\bf 65} (2002) 075002
  [arXiv:hep-ph/0104225];
  C.~H.~Chen and C.~Q.~Geng,
  ``The muon anomalous magnetic moment from a generic charged Higgs with
  SUSY,''
  Phys.\ Lett.\  B {\bf 511} (2001) 77
  [arXiv:hep-ph/0104151].


\bibitem{Stockinger:2006zn}
  For a recent review, see, for example,
  D.~Stockinger,
  ``The muon magnetic moment and supersymmetry,''
  J.\ Phys.\ G {\bf 34} (2007) R45
  [arXiv:hep-ph/0609168].

\bibitem{Pospelov:2005pr}
  M.~Pospelov and A.~Ritz,
  ``Electric dipole moments as probes of new physics,''
  Annals Phys.\  {\bf 318} (2005) 119
  [arXiv:hep-ph/0504231].

\bibitem{RamseyMusolf:2006vr}
  M.~J.~Ramsey-Musolf and S.~Su,
  ``Low energy precision test of supersymmetry,''
  Phys.\ Rept.\  {\bf 456} (2008) 1
  [arXiv:hep-ph/0612057].

\bibitem{Ibrahim:2007fb}
  T.~Ibrahim and P.~Nath,
  ``CP violation from standard model to strings,''
  Rev.\ Mod.\ Phys.\  {\bf 80} (2008) 577
  [arXiv:0705.2008 [hep-ph]].

\bibitem{Ellis:2008zy}
  J.~R.~Ellis, J.~S.~Lee and A.~Pilaftsis,
  ``Electric Dipole Moments in the MSSM Reloaded,''
  JHEP {\bf 0810} (2008) 049
  [arXiv:0808.1819 [hep-ph]].


\bibitem{Bennett:2008dy}
  G.~W.~Bennett {\it et al.},
  ``An Improved Limit on the Muon Electric Dipole Moment,''
  arXiv:0811.1207 [hep-ex].

\bibitem{Regan:2002ta}
  B.~C.~Regan, E.~D.~Commins, C.~J.~Schmidt and D.~DeMille,
  ``New limit on the electron electric dipole moment,''
  Phys.\ Rev.\ Lett.\  {\bf 88} (2002) 071805.

\bibitem{Baker:2006ts}
  C.~A.~Baker {\it et al.},
  ``An improved experimental limit on the electric dipole moment of the
  neutron,''
  Phys.\ Rev.\ Lett.\  {\bf 97} (2006) 131801.

\bibitem{Romalis:2000mg}
  M.~V.~Romalis, W.~C.~Griffith and E.~N.~Fortson,
  ``A new limit on the permanent electric dipole moment of Hg-199,''
  Phys.\ Rev.\ Lett.\  {\bf 86} (2001) 2505.


\bibitem{Semertzidis:1999kv}
  Y.~K.~Semertzidis {\it et al.},
  ``Sensitive search for a permanent muon electric dipole moment,''
  arXiv:hep-ph/0012087.

\bibitem{Ibrahim:2001ym}
  T.~Ibrahim, U.~Chattopadhyay and P.~Nath,
  ``Constraints on explicit CP violation from the Brookhaven muon g-2
  experiment,''
  Phys.\ Rev.\  D {\bf 64} (2001) 016010
  [arXiv:hep-ph/0102324].



\bibitem{cpsuperh}
  J.~S.~Lee, A.~Pilaftsis, M.~Carena, S.~Y.~Choi, M.~Drees, J.~R.~Ellis and C.~E.~M.~Wagner,
  Comput.\ Phys.\ Commun.\  {\bf 156} (2004) 283
  [arXiv:hep-ph/0307377];
  J.~S.~Lee, M.~Carena, J.~Ellis, A.~Pilaftsis and C.~E.~M.~Wagner,
  arXiv:0712.2360 [hep-ph].

\bibitem{Guasch:2001wv}
  J.~Guasch, W.~Hollik and S.~Penaranda,
  ``Distinguishing Higgs models in H $\to$ b anti-b / H $\to$ tau+ tau-,''
  Phys.\ Lett.\  B {\bf 515} (2001) 367
  [arXiv:hep-ph/0106027].

\bibitem{Carena:1998gk}
  M.~S.~Carena, S.~Mrenna and C.~E.~M.~Wagner,
  ``MSSM Higgs boson phenomenology at the Tevatron collider,''
  Phys.\ Rev.\  D {\bf 60} (1999) 075010
  [arXiv:hep-ph/9808312].

\bibitem{Ibrahim:2003tq}
  T.~Ibrahim and P.~Nath,
  ``Weak isospin violations in charged and neutral Higgs couplings from  SUSY
  loop corrections,''
  Phys.\ Rev.\  D {\bf 69} (2004) 075001
  [arXiv:hep-ph/0311242].

\bibitem{Ellis:2004fs}
  J.~R.~Ellis, J.~S.~Lee and A.~Pilaftsis,
  ``LHC signatures of resonant CP violation in a minimal supersymmetric  Higgs
  sector,''
  Phys.\ Rev.\  D {\bf 70} (2004) 075010
  [arXiv:hep-ph/0404167].


\bibitem{Carena:2000ks}
  M.~Carena, J.~R.~Ellis, A.~Pilaftsis and C.~E.~M.~Wagner,
  ``CP-violating MSSM Higgs bosons in the light of LEP 2,''
  Phys.\ Lett.\  B {\bf 495} (2000) 155
  [arXiv:hep-ph/0009212].

\bibitem{Schael:2006cr}
  S.~Schael {\it et al.}  [ALEPH Collaboration and DELPHI Collaboration and
                  L3 Collaboration and ],
  ``Search for neutral MSSM Higgs bosons at LEP,''
  Eur.\ Phys.\ J.\  C {\bf 47} (2006) 547
  [arXiv:hep-ex/0602042].

\bibitem{upsilon_visible}
  P.~Franzini {\it et al.},
  ``LIMITS ON HIGGS BOSONS, SCALAR QUARKONIA, AND ETA (B)'S FROM RADIATIVE
  UPSILON DECAYS,''
  Phys.\ Rev.\  D {\bf 35} (1987) 2883.
  J.~S.~Lee and S.~Scopel,
  ``Lightest Higgs boson and relic neutralino in the MSSM with CP violation,''
  Phys.\ Rev.\  D {\bf 75} (2007) 075001
  [arXiv:hep-ph/0701221].




\bibitem{Pilaftsis:1999qt}
  See, for example, A.~Pilaftsis and C.~E.~M.~Wagner,
  ``Higgs bosons in the minimal supersymmetric standard model with explicit  CP
  violation,''
  Nucl.\ Phys.\  B {\bf 553} (1999) 3
  [arXiv:hep-ph/9902371].

\end{thebibliography}
\end{document}